\documentclass[]{jfm}
\usepackage{siunitx}
\usepackage[font=small,labelfont=bf,justification=justified]{caption}
\usepackage{wrapfig}
\usepackage{graphicx}
\usepackage{newtxtext}
\usepackage{newtxmath}
\usepackage{natbib}
\usepackage[export]{adjustbox}
\usepackage{hyperref}
\hypersetup{
    colorlinks = true,
    urlcolor   = blue,
    citecolor  = black,
}
\usepackage{soul}
\usepackage{xcolor}
\usepackage{caption}
\usepackage{float}
\usepackage[labelsep=period]{caption}

\usepackage{hyperref}
\usepackage{amsmath}
\usepackage{epstopdf}
\usepackage{graphicx}
\usepackage{color}
\usepackage{soul}
\usepackage{setspace} 
\usepackage{amssymb}
\usepackage{amsmath}
\usepackage{multicol}
\usepackage{latexsym}
\usepackage{subfigure}
\usepackage{multirow}
\usepackage{setspace}

\usepackage{multirow}
\usepackage{arydshln}
\usepackage{amsmath,amssymb}
\usepackage{ragged2e}
\usepackage{tikz}
\usetikzlibrary{calc,topaths}
\usepackage{notoccite}

\usepackage{setspace}

\newcommand{\RomanNumeralCaps}[1]
\linenumbers
\renewcommand{\figurename}{Figure}

\title{\justify{Separate and joint clustering characteristics of large Stokes number sprays subjected to moderately turbulent co-flows}}

\author{Ali Rostami, Ri Li,
 \and Sina Kheirkhah\corresp{\email{sina.kheirkhah@ubc.ca}}}

\affiliation{\aff{}School of Engineering, University of British Columbia, Kelowna, British Columbia, Canada, V1V1V7}

\begin{document}
\maketitle

\begin{abstract}
Separate and joint droplets, clusters, and voids characteristics of sprays injected in a turbulent co-flow are investigated experimentally. Simultaneous Mie scattering and Interferometric Laser Imaging for Droplet Sizing along with separate hotwire anemometry are performed. The turbulent co-flow characteristics are adjusted using zero, one, or two perforated plates. The Taylor length scale-based Reynolds number varies from 10 to 38, and the Stokes number estimated based on the Kolmogorov time scale varies from 3 to 25. The results show that the mean length scale of the clusters normalized by the Kolmogorov length scale varies linearly with the Stokes number. However, the mean void length scale is on the order of the integral length scale. It is shown that the number density of the droplets inside the clusters is about 10 times larger than that in the voids. The ratios of the droplets number densities in the clusters and voids to the total number density are independent of the test conditions and equal 8.3 and 0.7, respectively. The joint probability density function of the droplets diameter and clusters area shows that the droplets with the most probable diameter are found in the majority of the clusters. It is argued that intensifying the turbulence broadens the range of turbulent eddy size in the co-flow which allows for accommodating droplets with a broad range of diameters in the clusters. The results are of significance for engineering applications that aim to modify the clustering characteristics of large Stokes number droplets sprayed into turbulent co-flows.
\end{abstract}


\section{Introduction}
\label{sec:Introduction}
Turbulent particle-laden flows are of relevance to many engineering applications such as gas turbine engine combustors used for aircraft propulsion, cyclones utilized for particle separation, industrial driers, and slurry pumps, see  \cite{Crowe2011Multiphase,m2016point,zhao2021numerical}. Thus, several theoretical, numerical, and experimental investigations have been carried out over the past decades, and many review papers have been published, see for example \cite{poelma2006particle,balachandar2010turbulent,brandt2022particle}. Review of these studies suggests that, among many non-dimensional parameters, the Stokes number (which is the ratio of particle to flow time scale, as defined in for example \cite{Crowe2011Multiphase}) primarily influences the characteristics of the turbulent particle-laden flows. Although past investigations are of significant importance as they provide insight into relatively small and moderate Stokes number flows ($St\lesssim 10$), majority of industrial applications relevant to turbulent particle-laden flows corresponds to relatively large Stokes numbers ($St \gtrsim 10$). The present study is motivated by the need for understanding the spray characteristics at relatively large Stokes numbers. As elaborated in \cite{baker2017coherent} and \cite{boddapati2020novel}, the interaction of the particles and the background turbulent flow leads to the formation of regions with relatively large and small number of particles, which are referred to as clusters and voids, respectively, and are relevant to the present study. A brief review of the literature related to the clusters and voids is provided below.

The inertial bias and sweep-stick mechanisms are proposed in the literature to elaborate the formation of clusters and voids in turbulent flows. The former suggests that during a particle and eddy interaction, the large eddies centrifuge out the particles and accumulate them in regions that feature small vorticities and large strain rates, as explained in \cite{maxey1987gravitational,squires1991preferential,wang1993settling}. Compared to the inertial bias mechanism, the sweep-stick mechanism suggests that the vorticity and strain rate fields may not be sufficient in explaining the positioning of the particles and hence formation of the clusters, see \cite{goto2008sweep}. The sweep-stick mechanism suggests that particles with $St \gtrsim 1$ tend to be positioned in the spatial locations with zero acceleration. Then, these particles are carried by large eddies in the flow, see \cite{goto2006self, monchaux2012analyzing, mora2021clustering,hassaini2022scale}.

Various tools and methods have been developed to identify the clusters and voids from the spatial distribution of the particles. For example, \cite{monchaux2010preferential}, \cite{tagawa2012three,frankel2016settling} utilized the Vorono\"{i} cells, \cite{andrade2022study} used the combined graph and Vorono\"{i} cells, \cite{fessler1994preferential} and \cite{villafane2016robust} implemented the box-counting method, and \cite{salazar2008experimental,saw2008inertial,sahu2016droplet} applied the radial distribution functions to identify the clusters. The Vorono\"{i} cells (relevant to the present investigation) can allow for estimating the degree of clustering, which is defined (see for example \cite{boddapati2020novel}) as the root mean square (RMS) of the Vorono\"{i} cells area ($\sigma$) divided by that if the droplets were distributed following a Random Poison Process (RPP), $\sigma_\mathrm{RPP}$, minus unity. Studies of \cite{reade2000effect, aliseda2002effect,yang2005two, monchaux2010preferential,obligado2014preferential,sumbekova2017preferential} show that the degree of clustering is influenced by three non-dimensional parameters that are the Stokes number ($St$) estimated based on the Kolmogorov time scale, Taylor length scale-based Reynolds number ($Re_\lambda$), and the droplets volume fraction ($\phi_\mathrm{v}$). \cite{reade2000effect, aliseda2002effect,yang2005two, monchaux2010preferential,obligado2014preferential} showed that the degree of clustering increases by increasing the Stokes number and maximizes at $St\approx 1$. \cite{sumbekova2017preferential} proposed a power-law correlation between the degree of clustering and $St$, $Re_\lambda$, and $\phi_\mathrm{v}$. Compared to \cite{reade2000effect, aliseda2002effect,yang2005two, monchaux2010preferential,obligado2014preferential}, the study of \cite{sumbekova2017preferential} showed that $St$ does not greatly influence the degree of clustering, however, this parameter scales with $Re_\lambda$ and the square root of $\phi_\mathrm{v}$. 

Few experimental studies, see \cite{zimmer2003simultaneous} and \cite{jedelsky2018air}, were carried out to quantify the position of the clusters and voids in turbulent spray flows. Their results show that the clusters are formed in the central region of the sprays, where small size droplets are located. Studies of \cite{sumbekova2017preferential,obligado2014preferential,monchaux2010preferential} showed that the Probability Density Function (PDF) of the cluster and void areas feature power-law correlations with the exponents of the power-law ranging from about -1.5 to -2.1 and -1.7 to -1.9, respectively. \cite{sumbekova2017preferential} showed that the PDFs of the cluster and void areas normalized by their corresponding mean value feature a power-law decay with an exponent of -5/3 for normalized cluster and void areas ranging from about 0.2 to 10.

In addition to the positions and the PDFs of the cluster and void areas, the length scale of the clusters and voids can be used to understand their relations with the length scale of the turbulent flow. The characteristic length scale of the clusters and voids are defined as the square root of the cluster and void mean areas, respectively. Studies of \cite{aliseda2002effect,obligado2014preferential, sumbekova2017preferential, sahu2016droplet,boddapati2020novel} showed that the cluster length scale is about 5--90 times the Kolmogorov length scale. For voids, \cite{sumbekova2017preferential} showed that the length scale can increase to about 200 times the Kolmogorov length scale. Power-law formulations were developed in \cite{sumbekova2017preferential}, and it was shown that the normalized cluster length scale is proportional to $St^{-0.25}Re_\lambda^{4.7}\phi_\mathrm{v}^{1.2}$. Also, \cite{sumbekova2017preferential} showed that, despite the Stokes number does not significantly influence the normalized voids length scale, this positively relates to $Re_\lambda$ and $\phi_\mathrm{v}$. 

Although the generated knowledge from the particle-laden flow studies (with a brief review presented above) is of significant importance, as it allows to understand the clustering characteristics of droplets corresponding to relatively small Stokes numbers ($St < 10$), many engineering applications feature droplets with large Stokes numbers. In such applications, it is of significant importance to understand the distribution of the droplet diameter and their number density within a given cluster and for $St > 10$. For example, for spray combustion-related applications (which feature Stokes numbers in excess of 10), the mean distance between the fuel droplets and their number density within a given cluster can influence the droplets evaporation rate as well as the flame location and the temperature distribution, see for example \cite{sahu2018interaction,hardalupas1994mass,akamatsu1996measurement,pandurangan2022spatial}. The objective of the present study is to investigate the effect of the Stokes number on both the geometrical (e.g. clusters and voids length scales) as well as the joint characteristics of the droplets and clusters/voids for $St > 10$. In the following, the methodology, data reduction, results, and concluding remarks are presented in Sections~\ref{sec:Experimental methodology}--\ref{Sec:Conclusions}, respectively.

\section{Experimental methodology}\label{sec:Experimental methodology}
The experimental setup, the utilized diagnostics, and the tested conditions are presented in this section.

\subsection{Experimental setup}\label{subsec:Experimentalsetup}
\begin{figure}
	\centerline{\includegraphics[width=0.95\textwidth]{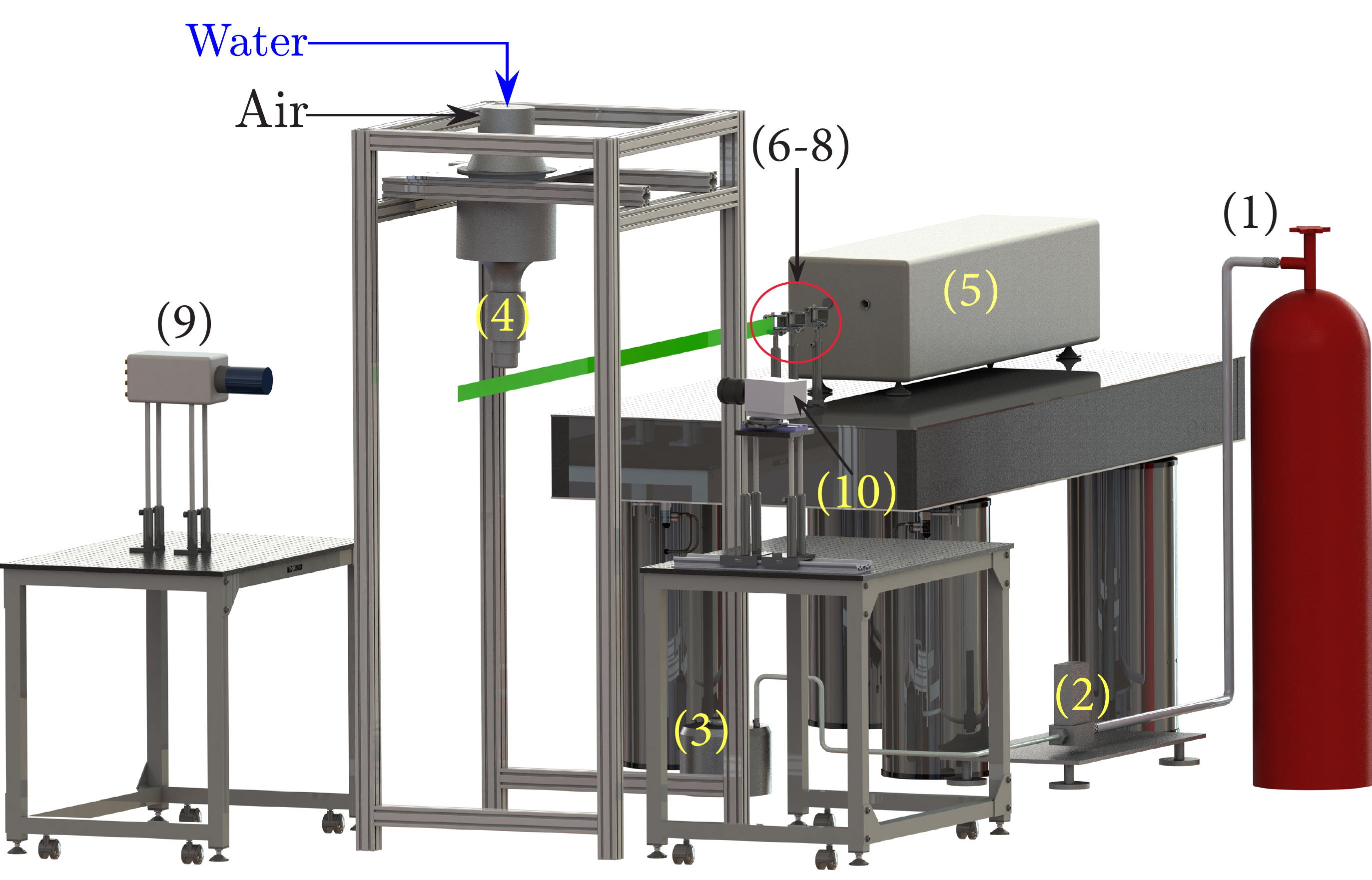}}
	\caption{The experimental setup. (1--3) are a nitrogen bottle, a pressure controller, and a pressurized water vessel. (4) is the nozzle section of the flow apparatus. (5) is a 532~nm Nd:YAG laser. (6--8) are the laser sheet forming optics and optomechanics. (9) and (10) are a Nova S12 camera and lens as well as a Zyla camera and lens for simultaneous shadowgraphy and ILIDS measurements.}
	\label{Fig:Setup}
\end{figure}

The experimental setup consists of a liquid delivery and flow apparatuses, which are shown in Fig.~\ref{Fig:Setup} as items (1--3) and item (4), respectively. A nitrogen bottle, item (1), in the figure as well as a dual-valve MCRH 2000 ALICAT pressure controller, item (2), were used to purge water inside a pressure vessel, item (3). Then, water flowed into the flow apparatus, see the blue arrow in Fig.~\ref{Fig:Setup}. In addition to water, a compressor was used to provide air into the flow apparatus, see the black arrow in Fig.~\ref{Fig:Setup}. The air flow rate was controlled using an MCRH 5000 ALICAT mass flow controller.

\begin{figure}
	\centerline{\includegraphics[width=1\textwidth]{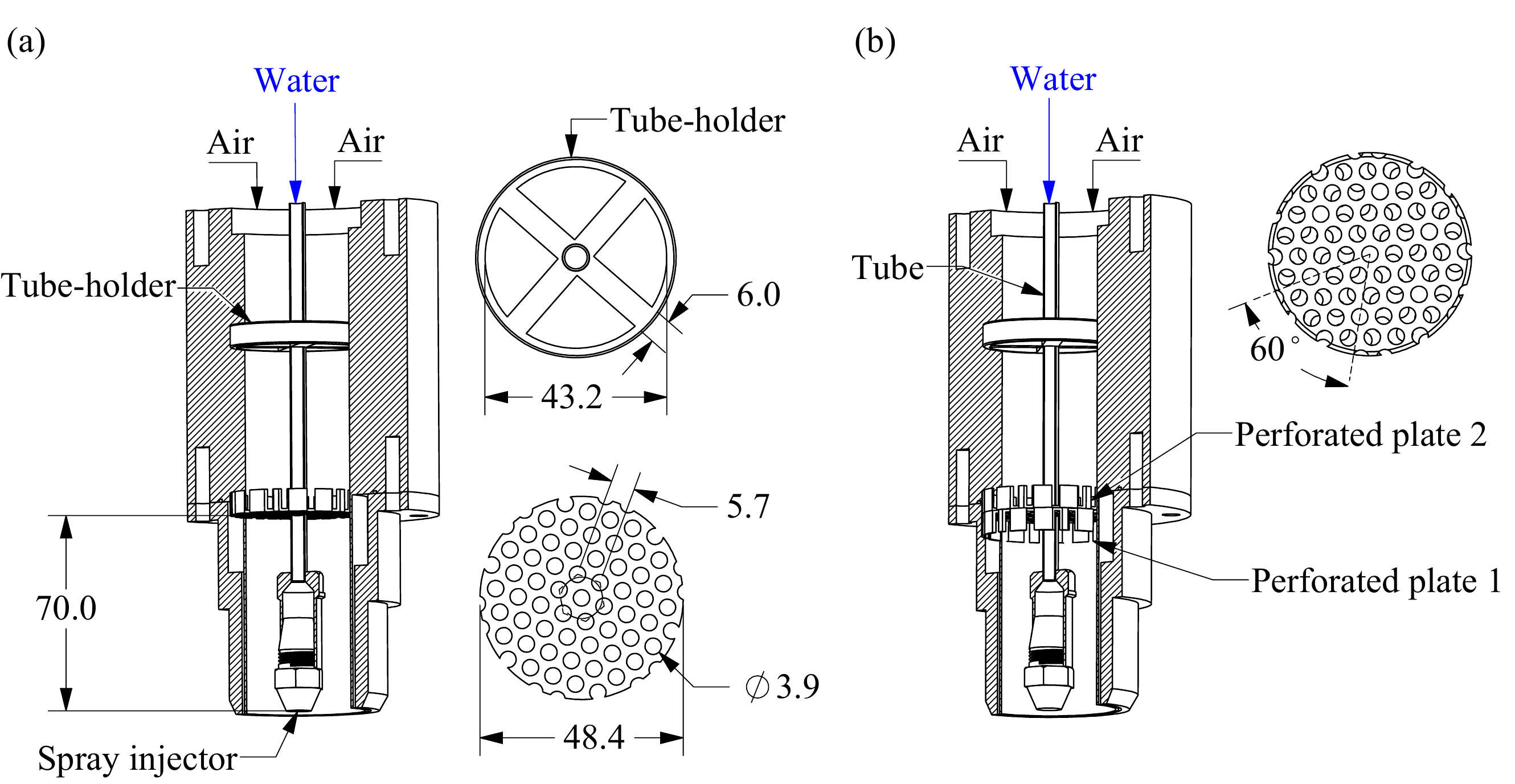}}
	\caption{(a) and (b) are the 3D drawings of the flow apparatus nozzle section for the second and third turbulence generation mechanisms, respectively.}
	\label{Fig:Nozzle geometry}
\end{figure}

The flow apparatus is composed of a diffuser section (with an area ratio of 4), a settling chamber (which is equipped with 5 equally spaced mesh screens), a contraction section (with an area ratio of 7), and a nozzle (with an inner diameter of 48.4~mm). Further details regarding the apparatus can be found in \cite{mohammadnejad2022new,kheirkhah2015consumption,kheirkhah2016experimental}. The nozzle section of the flow apparatus is shown in Fig.~\ref{Fig:Nozzle geometry}. In the present study, this section was modified to allow for producing a spray subjected to turbulent co-flow of air. The nozzle includes a 6.4~mm outer diameter tube (which carries water), a spray injector, a turbulence generation mechanism, and a ring-shaped tube-holder, which was press-fit against the inner wall of the nozzle and supported the tube vertically using 4 bars and a collar. The injector was a pressure swirl atomizer from Delavan (model 6330609), which produced a polydisperse spray. The spray flow rate depends on the vessel pressure, and separate calibration experiments were performed to obtain the relation between the vessel pressure and the spray flow rate, as discussed in Appendix~A.

Three turbulence generation mechanisms were used in the present study. Either no perforated plate, one perforate plate (see Fig.~\ref{Fig:Nozzle geometry}(a)), or two perforated plates attached back-to-back (see Fig.~\ref{Fig:Nozzle geometry}(b)) were utilized. The outer diameter of each perforated plates is 48.4~mm, matching the inner diameter of the nozzle. Each perforated plate features 3.9~mm diameter holes arranged in a hexagonal pattern. For the first turbulence generation mechanism, the perforated plates in Fig.~\ref{Fig:Nozzle geometry} were removed. For the third turbulence generation mechanism, the plates were rotated by 60$^\mathrm{o}$ with respect to one another, similar to \cite{kheirkhah2014topology}. For the second turbulence generation mechanism, the planes containing the holes; and, for the third turbulence generation mechanism, the center plane of the two perforated plates were positioned 76.0~mm upstream of the nozzle exit plane, similar to the studies of \cite{kheirkhah2015consumption} and \cite{kheirkhah2016experimental}.

\subsection{Diagnostics}\label{subsec:Diagnostics}
Separate hotwire anemometry (HWA) as well as simultaneous Mie scattering and Interferometric Laser Imaging for Droplet Sizing (ILIDS) were performed. The hotwire anemometry was used to characterize the background turbulent flow. A schematic of the diagnostics used in the present study is shown in Fig.~\ref{Fig:Setup}. The hardware configuration for the HWA is identical to that used in \cite{mohammadnejad2022new}; and, a separate illustration is not presented in Fig.~\ref{Fig:Setup}. For all test conditions, the HWA data was acquired at a frequency of 100~kHz and for 90~s, corresponding to 9,000,000 data points. The simultaneous Mie scattering and ILIDS images were collected at a frequency of 10~Hz and for a duration of 100~s, corresponding to 1000 image pairs. Further details regarding the HWA, Mie scattering, and ILIDS diagnostics are provided in the following.

\subsubsection{Hotwire anemometry}\label{subsubsec:HWA}
The hardware of the HWA system consists of a probe (model 55P01 from Dantec), a probe support (model 9055H0261 from Dantec), and two motorized translational stages (MTS50-Z8 from Thorlabs). The probe is a single wire sensor, which is 3~mm long (with an active sensor length of 1.25~mm) and has a diameter of 5~$\mu$m. A mini-constant temperature anemometry (mini-CTA) circuit (model 9054T0421 from Dantec) maintains the wire temperature, with an overheat ratio of 0.7. The motorized translational stages featured a 50~mm range of operation, which was sufficient for the present study. Further details regarding the HWA system as well as the calibration procedure can be found in \cite{mohammadnejad2022new}.

\begin{figure}
\centerline{\includegraphics[width=0.6\textwidth]{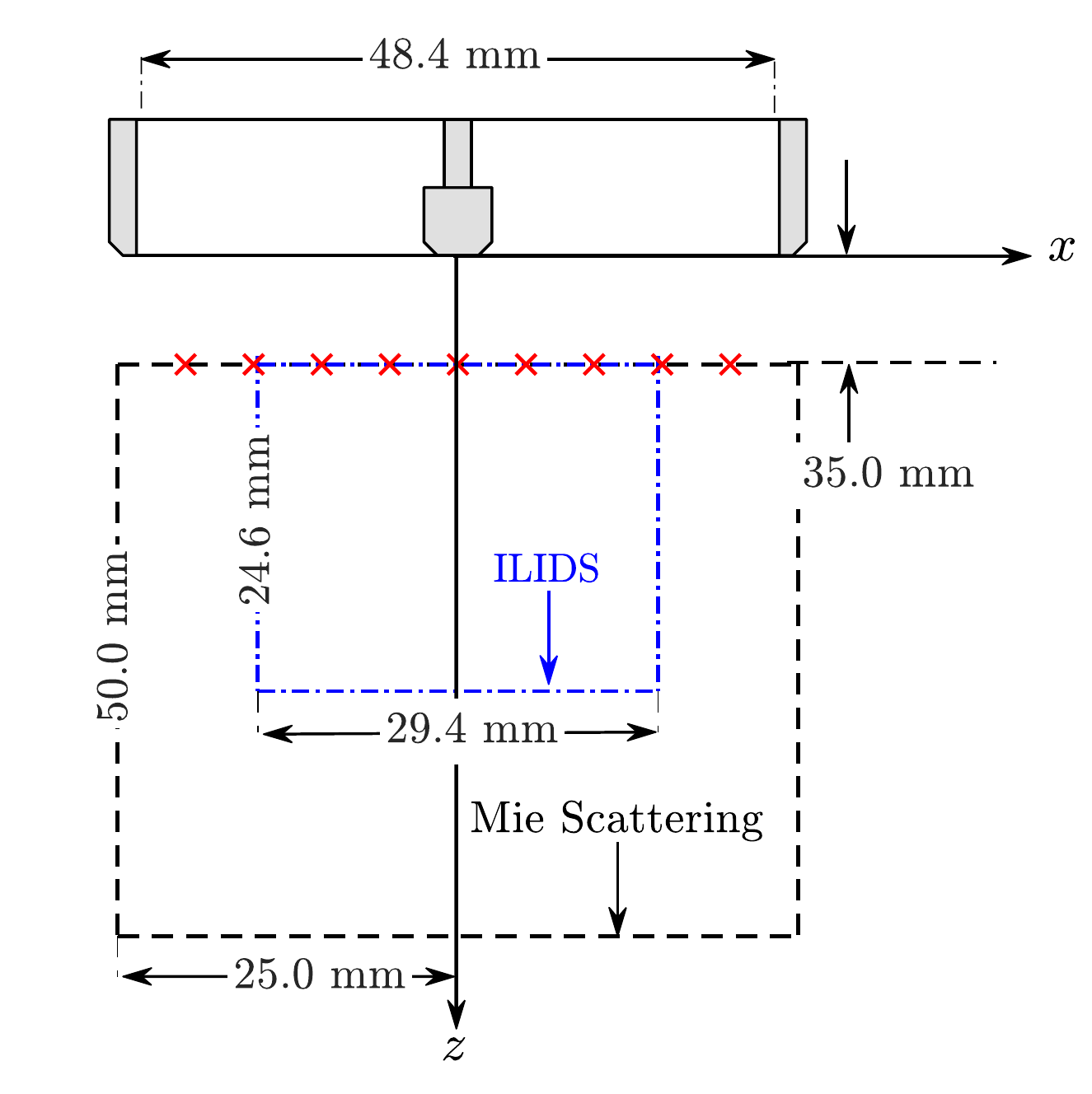}}
	\caption{The coordinate system and the measurements locations. The red cross data points present the locations at which the hotwire anemometry was performed. The dashed black and dotted dashed blue squares are the regions of interest for the Mie scattering and ILIDS measurements, respectively. The minimum vertical distance for the above measurements is 35.0~mm from the nozzle exit plane, and this distance shown in the figure is not to scale.}
	\label{Fig: Coordinate system}
\end{figure}

A Cartesian coordinate system was used in the present investigation. The origin of the coordinate system is at the exit plane of the nozzle section and at the nozzle centerline, as shown in Fig.~\ref{Fig: Coordinate system}. The $z$--axis coincides with the nozzle centerline. The $x$--axis is normal to the $z$--axis and is parallel with the laser sheet shown in Fig.~\ref{Fig:Setup}. The HWA was performed at $z = 35.0$~mm and at horizontal locations spaced by 5.0~mm along the $x$--axis, ranging from $x = -20.0$~mm to 20.0~mm as shown as the red cross data symbols in Fig.~\ref{Fig: Coordinate system}. 

\subsubsection{Mie scattering}\label{subsubsec:Miescattering}
The Mie Scattering hardware consists of a pulsed Nd:YAG laser~(Lab-Series-170 from Spectra Physics, shown by item~(5) in Fig.~\ref{Fig:Setup}), the sheet forming optics (see, items~(6--8)), and a camera equipped with collection optics (item~(9)). The laser produces a 1064~nm beam, which is converted to a 532~nm beam using a harmonic generator. The laser was operated at a reduced (compared to its maximum) but fixed energy in order to avoid saturation in the collected Mie scattering images. The laser beam was 8.0~mm in diameter, which was converted to a 40~mm high and 1~mm thick laser sheet using a plano-concave cylindrical lens with a focal length of -100~mm, a plano-convex cylindrical lens with a focal length of 500~mm, and a cylindrical lens with a focal length of 1000~mm, see items~(6--8) in Fig.~\ref{Fig:Setup}. The centerline of the collimated laser sheet was positioned at $z=55.0$~mm. The Mie scattering images were acquired using a Photron Fastcam Nova S12 camera equipped with a Macro Sigma lens, which had a focal length of $f =105$~mm and its aperture size was set to $f/2.8$. A bandpass filter with a center wavelength and Full Width at Half Maximum of 532 and 20~nm was mounted on the camera lens. By adjusting the working distance of the camera, a field of view of 70.0~mm along $x$--axis and 70.0~mm along the $z$--axis was obtained, which corresponds to a pixel resolution of $70~\mathrm{mm}/1024~\mathrm{pixels}=68~\mu\mathrm{m}/\mathrm{pixel}$. For analysis and presentation purposes, the above field of view was cropped to a 50.0~mm$\times$50.0~mm square, which is shown by the black dashed box in Fig.~\ref{Fig: Coordinate system}.

\subsubsection{Interferometric Laser Imaging for Droplet Sizing}\label{subsubsec:ILIDSdiagnostics}
The Interferometric Laser Imaging for Droplet Sizing was performed to measure the droplets diameter, similar to~\cite{qieni2016high,bocanegra2015measuring,garcia2021interferometric}. In ILIDS, the droplet diameter is measured by analyzing the interference pattern of the reflected and first-order refracted rays scattered from spherical droplets that are illuminated by a laser. The ILIDS hardware includes the laser and the laser-sheet forming optics, see items (5--8) in Fig.~\ref{Fig:Setup}, which are identical to those used for the Mie scattering technique as well as an sCMOS camera equipped with a Macro sigma lens ($f = 105$~mm and aperture size of $f/2.8$) and a high-precision rotary stage, see item (10). The camera is a 5.5 Zyla from Andore, which has a 2560~pixels$\times$2160~pixels sensor. The ILIDS field of view was 29.4~mm along the $x$--axis and 24.6~mm along the $z$-axis, which led to a pixel resolution of $29.4~\mathrm{mm}/2560~\mathrm{pixels} = 11.4~\mu\mathrm{m/pixel}$. Using the rotary stage, the orientation of the camera and lens was adjusted with respect to the direction of the laser sheet. Specifically, the angle between the lens centerline and the laser sheet, referred to as $\theta$, was set to 69$^\mathrm{o}$ as per the recommendation of \cite{sahu2016droplet}. It was confirmed this angle provided the best clarity of the fringe patterns. Finally, the formulation in \cite{hayashi49measurements} and \cite{thimothee2016experimental} was used to estimate the droplet diameter, which is given by

 \begin{equation}\label{eq:dropletdiameter}
 	d=\dfrac{2\lambda_\mathrm{L} N}{\alpha}\left[\cos\left(\frac{\theta}{2}\right)+\dfrac{m\sin(\frac{\theta}{2})}{\sqrt{m^{2}-2m\cos(\frac{\theta}{2})+1}})\right]^{-1},
 \end{equation}
where $\lambda_\mathrm{L}$ is the wavelength of the laser and is 532~nm. In Eq.~(\ref{eq:dropletdiameter}), $m$ is the droplets index of refraction, which equals 1.33 for water, and $\alpha$ is the collection angle of the scattered light. This angle depends on the utilized lens diameter ($d_\mathrm{l}$) and the distance between the camera lens and the laser sheet ($c$). The collection angle is calculated using $\alpha=2\arctan[d_\mathrm{l}/(2c)]$. In Eq.~(\ref{eq:dropletdiameter}), $N$ is the number of the fringe patterns and is estimated using the procedure discussed in the next section. Since the viewing angle of the ILIDS camera is different than that of the Mie scattering camera, a discrepancy exists in estimating the location of the droplets using the Mie scattering and ILIDS techniques. Such discrepancy needs to be addressed for estimating the joint characteristics of the droplets and clusters/voids. This discrepancy was addressed by registering the Mie scattering images onto the ILIDS images, with details discussed in Appendix~B.

\subsection{Test conditions}\label{subsec:Testconditions}
In total, 13 experimental conditions were tested, with the corresponding details tabulated in Table~\ref{tab:tested onditions}. The first row in the table highlights the test condition for which no co-flow is utilized and the spray is issued into the quiescent air. The first column in the table presents the utilized turbulence generation (TG) mechanism, with 0PP, 1PP, and 2PP referring to zero, one, and two perforated plates, respectively. For each turbulence generation mechanism, the mean bulk flow velocities of 3.5, 7.0, 10.5, and 14.0~m/s were tested. The root-mean-sqaure (RMS) of the streamwise velocity ($u'_0$) and the integral length scale ($\Lambda$) estimated at $x=0$~mm and $z = 35.0$~mm are listed in the third and fourth columns of Table~\ref{tab:tested onditions}, respectively. The integral length scale was calculated using the Taylor's frozen turbulence hypothesis, see \cite{taylor1938spectrum}, and following the procedure detailed in \cite{mohammadnejad2022new}. Specifically, the integral length scale was calculated as the multiplication of the mean bulk flow velocity and the integral of the streamwise velocity auto-correlation from $t =0$ to the first time that the auto-correlation becomes zero. The Taylor ($\lambda$) and Kolmogorov ($\eta$) length scales were calculated using $\lambda = \Lambda(u^\prime_0 \Lambda/\nu)^{-0.5}$ and $\eta = \Lambda(u^\prime_0 \Lambda/\nu)^{-0.75}$, with $\nu$ being the air kinematic viscosity estimated at the laboratory temperature. The values of $\lambda$ and $\eta$ are tabulated in the fifth and sixth columns of Table~\ref{tab:tested onditions}. For all test conditions, the most probable droplet diameters ($d^*$) were obtained using the ILIDS diagnostic and are listed in the seventh column of the table. Further details regarding the size distribution of the droplets are discussed in subsection~\ref{subsec:Resultsbackgroundflow}.

The Taylor length scale-based Reynolds number, the Stokes number, and the liquid volume fraction are non-dimensional parameters that can potentially influence the interaction of the droplets with the background turbulent flow. The Taylor length scale based-Reynolds number was estimated using $Re_\lambda = u^\prime_0 \lambda/\nu$; and, henceforth, for brevity, $Re_\lambda$ is referred to as the Reynolds number, with the corresponding values listed in the eighth column of Tabel~\ref{tab:tested onditions}. In the present study, $Re_\lambda$ varies from about 10 to 36, which corresponds to relatively moderate values. Following \cite{reade2000effect}, the Stokes number was calculated using
\begin{equation}\label{Eq:St}
St=\frac{1}{18}\frac{\rho_\mathrm{W}}{\rho_\mathrm{A}}(\frac{d}{\eta})^2
\end{equation}
where $\rho_\mathrm{W}$ and $\rho_\mathrm{A}$ are the water and air densities, respectively, both estimated at the laboratory temperature. In Eq.~(\ref{Eq:St}), $d$ is the droplet diameter. In  the present study, $d$ was taken as the most probable droplet diameter for estimation of the Stokes number. The values of $St$ are tabulated in Table~\ref{tab:tested onditions} and range from about 3 to 25, which are relatively large compared to those of the studies that were performed in multi-phase wind tunnels. Following the definition provided in \cite{elghobashi1994predicting}, the liquid volume fraction was calculated from
\begin{equation}
\label{eq:phiv}
\phi_\mathrm{v} = n\frac{V_\mathrm{d}}{V_\mathrm{ROI}},
\end{equation}
where $n$ is the mean number of droplets within the volume of the region of interest, $V_\mathrm{ROI}$. In Eq.~(\ref{eq:phiv}), $V_\mathrm{d}$ is the droplet volume. For each test condition, $n$ was estimated using the Mie scattering technique and averaged over all collected Mie scattering images. $V_\mathrm{ROI} = 50\times50\times1~\mathrm{mm}^3$ and $V_\mathrm{d} = (\pi/6)d^{*3}$. The values of $\phi_\mathrm{v}$ are listed in the last column of Table~\ref{tab:tested onditions} and change from about $1\times10^{-6}$ to $2\times10^{-6}$. Following \cite{elghobashi1994predicting}, the estimated liquid volume fractions are relatively small, rendering the tested sprays as dilute. In essence, compared to past studies, the non-dimensional parameters of the present study correspond to dilute sprays with moderate Reynolds but large Stokes numbers.
 
\begin{table}
	\begin{center}
	\def~{\hphantom{0}}
 \scalebox{1.15}{
	\begin{tabular}{l c c c c c c c c c c}
		
 TG & $U$ (m/s) & $u^\prime_0\mathrm{(m/s)}$ & $\Lambda$ (mm) & $\lambda$ (mm) & $\eta (\mu \mathrm{m}$)& $d^*(\mu \mathrm{m}$ ) & $Re_\lambda$ &  $St$ & $\phi_\mathrm{v} (\mathrm{x 10^6})$ \\
         N.A.  &	0.0  &  N.A.   &	N.A.  &  N.A.   &	N.A.   &	34  &   N.A.   & N.A.  &	1.8 \\
        \textcolor{blue}{0PP}  &	3.5  &	0.57  &	5.9	 &  0.39  &	102  &	29  &	14.9  &	4.5  &	1.8 \\
         \textcolor{blue}{0PP}  &	7 .0 &	1.07  &	5.8  &  0.29  &	63  &	29  &	20.3  &	11.4 &	1.2 \\
          \textcolor{blue}{0PP}  & 10.5  &	1.55  &	7.1  &  0.26  &	50  &	29  &	27.1  &	18.5 &	1.0 \\
          \textcolor{blue}{0PP}  &	14   &	2.14  &	10.0 &	0.26  &	43  &	29  &	37.8  &	25.3 &	1.1 \\
          \textcolor{green}{1PP}  &	3.5  &	0.4   &	4.0  &	0.39  &	121  &	29  &	10.3  &	3.2  &	2.1 \\
         \textcolor{green}{1PP}  &	7.0  &	0.85  &	4.7	 &  0.29  &	71  &	29  &	16.3  &	9.2  &	1.6 \\
         \textcolor{green}{1PP}  &	10.5 &	1.22  &	4.7  &	0.24  &	54  &	29  &	19.5  &	15.9 &	1.4 \\
         \textcolor{green}{1PP}  &	14.0 &	1.63  &	5.0	 &  0.21  &	44  &	28  &	23.3  &	21.4 &	1.2 \\
         \textcolor{red}{2PP}  &	3.5  &	0.64  &	4.9  &	0.34  &	89  &	29  &	14.5  &	5.7	 &2.2 \\
         \textcolor{red}{2PP}	 & 7.0   &	1.35  &	5.9  &	0.26  &53  &	29  &	23.0  &	16.5 &	1.6\\
         \textcolor{red}{2PP}  &	10.5 &	1.55  &	7.3  &	0.27  &	51  &	29  &	27.5  &	18.3 &	1.3 \\
        \textcolor{red}{2PP}  &	14.0 &	1.93  &	10.0 &	0.28  &	47  &	28  &	35.9  &	20.2 &	1.2 \\
         
	\end{tabular}
 }
    \caption{Test conditions.}
     \label{tab:tested onditions}
   \end{center}
\end {table}

\begin{figure}
	\centerline{\includegraphics[width=1\textwidth]{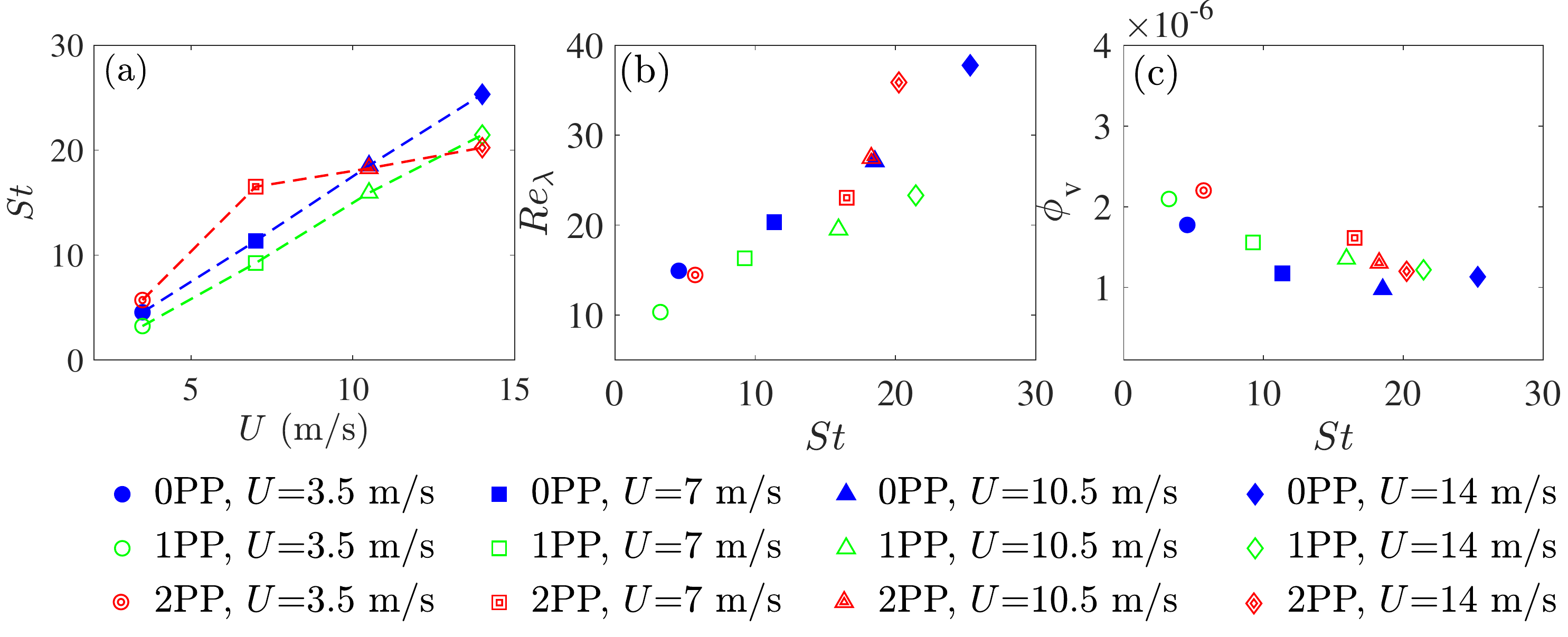}}
	\caption{Variations of (a) the Stokes number versus mean bulk flow velocity, (b) the Taylor length scale-based Reynolds number versus Stokes number, and (c) the liquid volume fraction versus the Stokes number. The blue, green, and red colors correspond to turbulence generation mechanisms with zero, one, and two perforated plates, respectively.}
	\label{Fig: tested condition}
\end{figure}

For all test conditions with the co-flow, $St$, $Re_\lambda$, and $\phi_\mathrm{v}$ vary by changing the mean bulk flow velocity and the turbulence generation mechanism. Variations of $St$ versus $U$, $Re_\lambda$ versus $St$, and $\phi_\mathrm{v}$ versus $St$ are presented in Figs.~\ref{Fig: tested condition}(a--c), respectively. As can be seen, the variations of these non-dimensional parameters are mostly influenced by the mean bulk flow velocity. That is increasing $U$ increases $St$ and $Re_\lambda$ but decreases $\phi_\mathrm{v}$. It is important to note that, in the present study, changing the turbulence generation mechanism for a fixed value of $U$ and changing $U$ for a given turbulence generation mechanism both change the background RMS velocity fluctuations, which changes $St$, $Re_\lambda$, and $\phi_\mathrm{v}$. In the following sections, when possible, comparisons are made at fixed values of Reynolds number but varying Stokes number and vice versa.

\section{Data Reduction}\label{sec:datareduction}
Of prime importance to the present study are identifications of clusters and voids as well as estimating the droplets diameter. The former and the later are obtained using the Mie scattering and the Interferometric Laser Imaging for Droplet Sizing, respectively, with details provided in the following subsections.

\subsection{Clusters and voids identification}\label{subsec:Miedatareduction}
\begin{figure}
	\centerline{\includegraphics[width=1\textwidth]{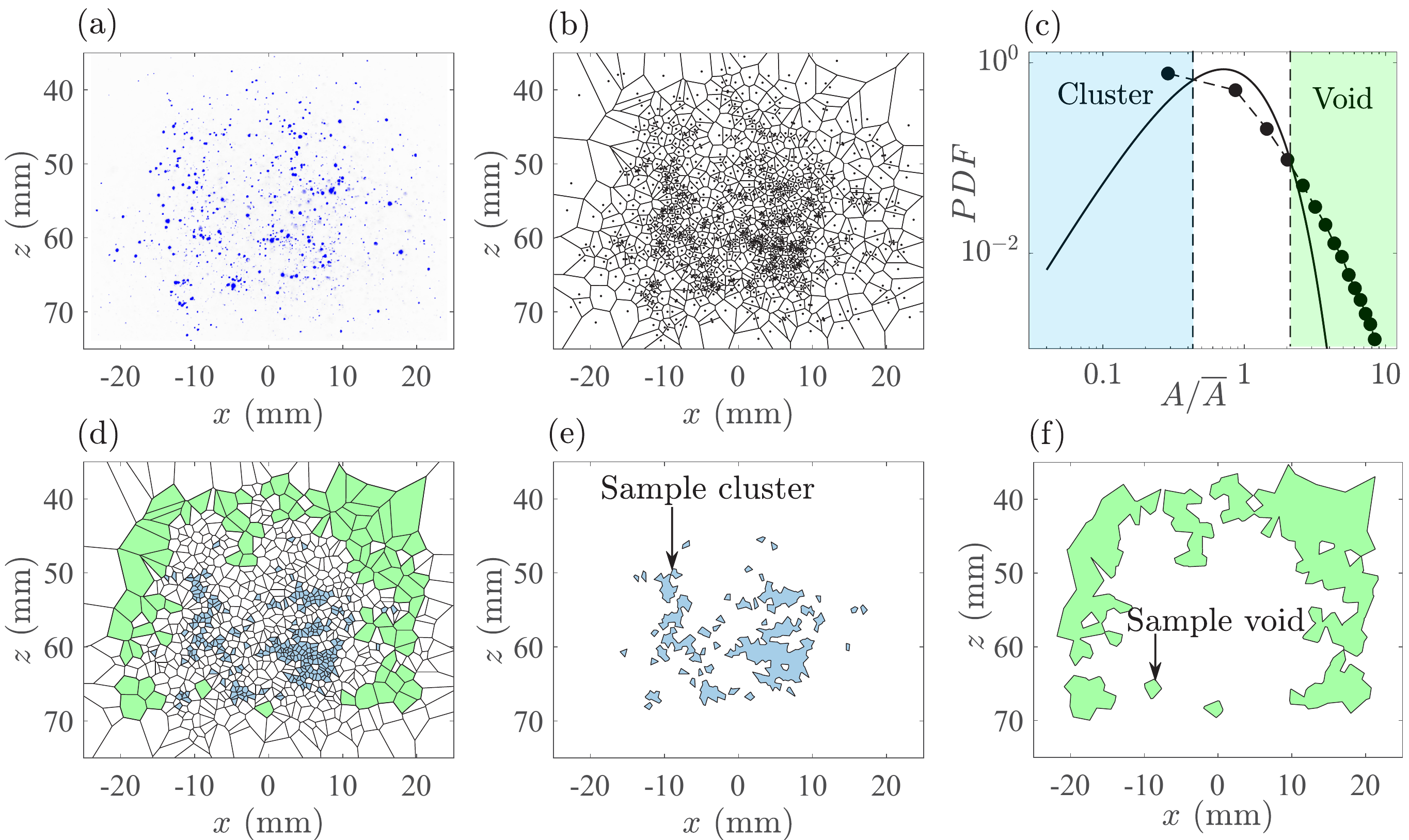}}
	\caption{(a) Representative raw Mie scattering image corresponding to the no co-flow test condition. (b) Centers of the droplets in (a) and the Vorono\"{i} cells. (c) PDF of the Vorono\"{i} cells areas normaliazed by their mean. Overlaid in (c) is the $PDF_\mathrm{RPP}$ from Eq.~(\ref{eq:random}). The dashed lines in (c) are $A/\overline{A} = 0.5$ and 2.2. (d) presents the cells with areas related to clusters (blue cells) and voids (green cells). (e) and (f) are the clusters and voids corresponding to the Mie scattering image in (a), respectively.}
	\label{Fig: Data Reduction}
\end{figure}

The procedures followed to identify the clusters and voids are illustrated in Fig.~\ref{Fig: Data Reduction}. A representative raw Mie scattering image corresponding to the no co-flow test condition is shown in Fig.~\ref{Fig: Data Reduction}(a). The results in Fig.~\ref{Fig: Data Reduction}(a) were binarized to identify the centers of the droplets, which are shown by the black circular data points in Fig.~\ref{Fig: Data Reduction}(b). It is important to note that the laser intensity features a nearly top-hat profile in the region of interest. Also, the Mie scattering background field (averaged over 1000 images taken with the camera lens capped) is about 0.2\% of the maximum acquired intensity. Thus, normalizing the Mie scattering images by the spatially varying laser intensity as well as subtracting the background field from the Mie scattering images did not influence the process for identifying the droplets centers in Fig.~\ref{Fig: Data Reduction}(b). Using the centers of the droplets along with the ``voronoi'' function in MATLAB, the Vorono\"{i} cells around each droplet were obtained and overlaid on Fig.~\ref{Fig: Data Reduction}(b) using the black lines.

The probability density function of the Vorono\"{i} cells area ($A$) normalized by their mean area ($\overline{A}$) was calculated using all Mie scattering images of  the no  co-flow test condition, and the resultant PDF is presented in Fig.~\ref{Fig: Data Reduction}(c) with the black circular data points. Provided the droplets spatial distribution followed the Random Poison Process (RPP), the PDF of $A/\overline{A}$, referred to as $PDF_\mathrm{RPP}$, could be estimated using that in \cite{ferenc2007size} which is given by

\begin{equation}
\label{eq:random}
PDF_\mathrm{RPP}=\frac{b^a}{\Gamma(a)}\left(A/\overline{A}\right)^{a-1}\exp^{(-b A/\overline{A})}.
\end{equation}
In Eq.~(\ref{eq:random}), $a$ and $b$ are fitting parameters with $a=b=3.5$, and $\Gamma$ is the gamma-function with $\Gamma(a) = 3.32$. The variation of $PDF_\mathrm{RPP}$ versus $A/\overline{A}$ was obtained and presented by the black solid curve in Fig.~\ref{Fig: Data Reduction}(c). Additionally, using MATLAB, 1000 images were synthetically generated by distributing particles inside the domain of investigation randomly, with the number of droplets identical to that shown in Fig.~\ref{Fig: Data Reduction}(a), and the PDF of $A/\overline{A}$ was calculated. It was confirmed that the PDF of the Vorono\"{i} cells normalized area for randomly distributed particles closely follows the right-hand-side of Eq.~(\ref{eq:random}). Next, the intersections of $PDF_\mathrm{RPP}$ and the $PDF(A/\overline{A})$ for the results in Fig.~\ref{Fig: Data Reduction} were obtained, which are shown by the dashed lines corresponding to $A/\overline{A} = 0.5$ and 2.2. The Vorono\"{i} cells with area smaller than $0.5 \overline{A}$ were labeled as clusters, and Vorono\"{i} cells with area larger than $2.2\overline{A}$ were labeled as voids. The identified clusters and voids corresponding to the Mie scattering image in Fig.~\ref{Fig: Data Reduction}(a) are shown by the blue and green color cells in Fig.~\ref{Fig: Data Reduction}(d). As can be seen, clusters or voids may feature connected boundaries forming larger clusters and voids. Following \cite{andrade2022study}, in the present study, the graph theory was used to identify and group the clusters/voids cells that are interconnected. For the results presented in Fig.~\ref{Fig: Data Reduction}(d), the identified clusters and voids are separately shown in Figs.~\ref{Fig: Data Reduction}(e) and (f), respectively. These clusters and voids were used for further analysis in Section~\ref{sec:Results}.

\subsection{Droplets location and diameter estimation}\label{subsec:ILIDSdatareduction}

\begin{figure}
	\centerline{\includegraphics[width=1\textwidth]{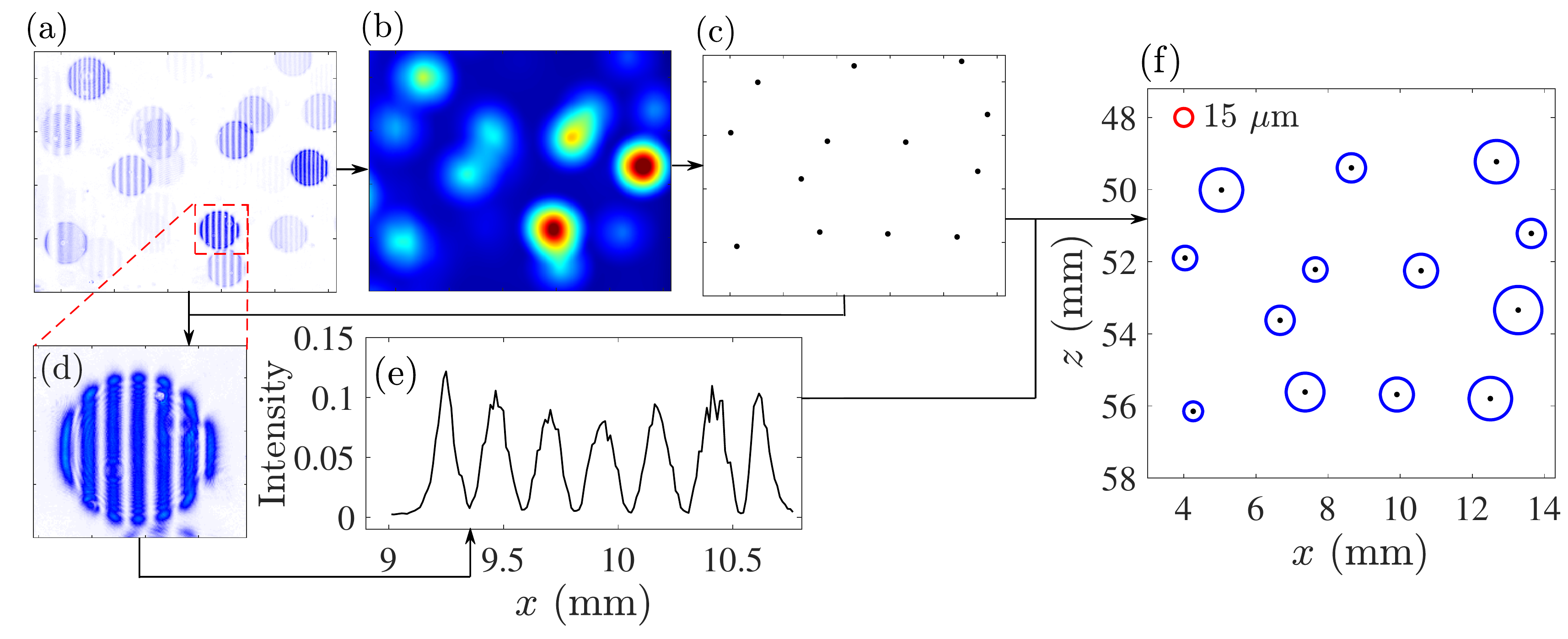}}
	\caption{(a) Representative raw ILIDS image corresponding to the test condition with two perforated plates and mean bulk flow velocity of 14.0~m/s. (b) is the convolution of the results in (a) using a disk-shaped mask. (c) is the identified droplets centers. (d) is the inset of (a), highlighting a sample fringe pattern. (e) shows the variation of the light intensity normal to the fringe pattern in (d). (f) presents the droplets centers and their corresponding diameters.}
	\label{Fig:ILIDS Data Reduction}
\end{figure}

The ILIDS images were reduced to estimate the droplets location and diameters. A summary of the processes followed to reduce the ILIDS images is illustrated in Fig.~\ref{Fig:ILIDS Data Reduction}. A representative raw ILIDS image is shown in Fig.~\ref{Fig:ILIDS Data Reduction}(a), which corresponds to the test condition with two perforated plates and the mean bulk flow velocity of 14.0~m/s. Following the procedure used in \cite{bocanegra2015measuring}, the image shown in Fig.~\ref{Fig:ILIDS Data Reduction}(a) was convoluted with a disk-shaped mask and the resultant image is shown in Fig.~\ref{Fig:ILIDS Data Reduction}(b). The convoluted image features local maxima, which correspond to the centers of the droplets. The locations of the droplets centers were obtained, with the corresponding results shown by the black circular data points in Fig.~\ref{Fig:ILIDS Data Reduction}(c). The variations of the light intensity along the lines that pass through the droplets centers and are normal to the corresponding fringe pattern were considered, with a sample fringe pattern and light intensity variation for one droplet shown in Figs.~\ref{Fig:ILIDS Data Reduction}(d)~and~\ref{Fig:ILIDS Data Reduction}(e), respectively. Then, the Fast Fourier Transform of the light intensity variation corresponding to each droplet was obtained and the number of fringes was calculated. Finally, the droplet diameter was calculated using the number of fringe patterns and Eq.~(\ref{eq:dropletdiameter}). Figure~\ref{Fig:ILIDS Data Reduction}(f) presents the centers of the droplets as well as the blue circles, with their diameter relating to the droplets' diameter. The diameters of the blue circles scale with the diameter of the red circle shown in the figure.

\section{Results}\label{sec:Results}
The results are grouped into four subsections. In the first subsection, the characteristics of the background turbulent flow and the droplets diameter are discussed. In the second subsection, the degree of clustering is investigated. In the third subsection, the geometrical characteristics of the clusters and voids are presented. Finally, the joint characteristics of the clusters/voids and the droplets are presented in the last subsection.

\subsection{The background flow and droplet diameter characteristics}\label{subsec:Resultsbackgroundflow}
\begin{figure}
	\centerline{\includegraphics[width=1\textwidth]{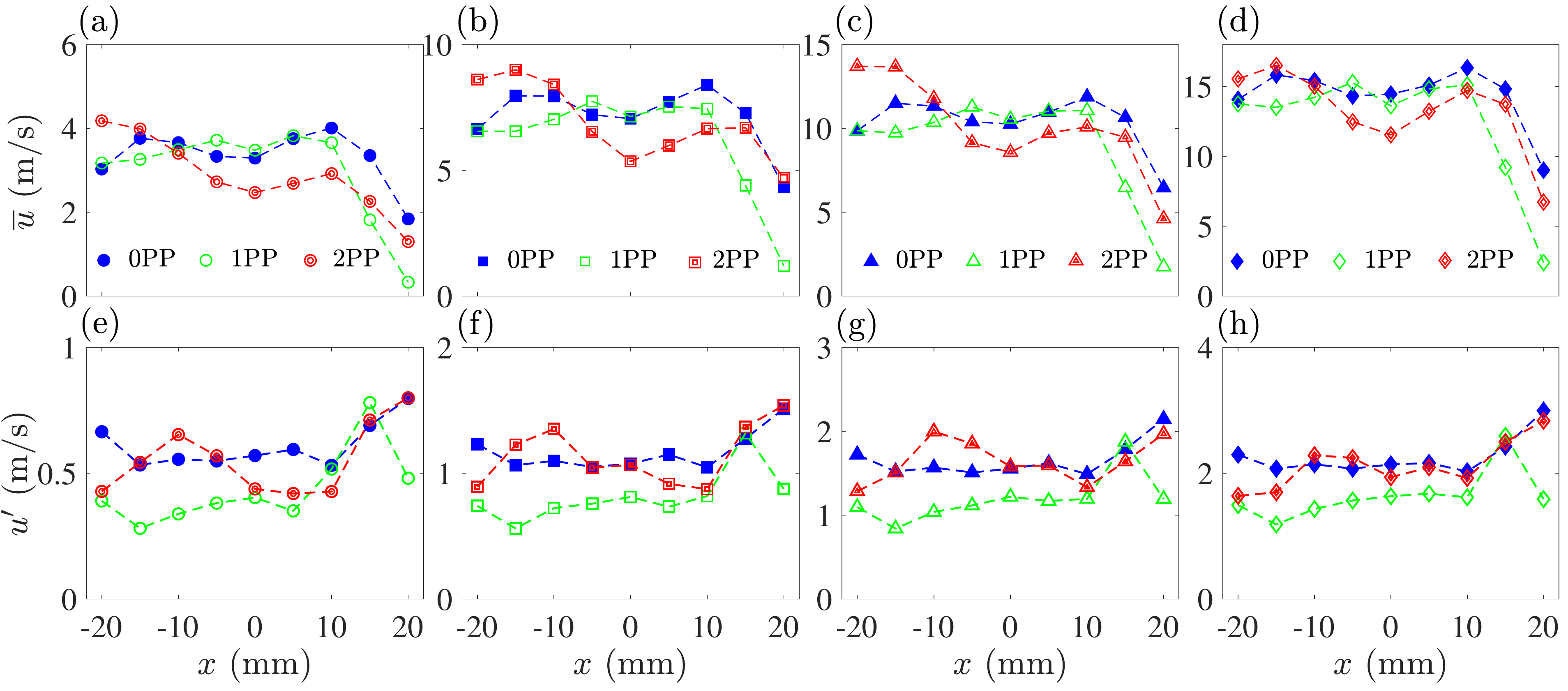}}
	\caption{(a--d) are the mean streamwise velocity for the mean bulk flow velocities of 3.5, 7.0, 10.5, and 14.0~m/s, respectively. (e--h) are the root-mean-square streamwise velocity fluctuations for the mean bulk flow velocities of 3.5, 7.0, 10.5, and 14.0~m/s, respectively.}
	\label{Fig: Velocity profile}
\end{figure}

The variations of the axial velocity mean ($\overline{u}$) and RMS fluctuations ($u^\prime$) along the $x$--axis are presented in the first and second rows of Fig.~\ref{Fig: Velocity profile}, respectively. The results in the first to fourth columns correspond to the mean bulk flow velocities of 3.5, 7.0, 10.5, and 14.0~m/s, using circular, square, triangular, and diamond-shaped data symbols, respectively. The blue, green, and red colors pertain to zero, one, and two perforated plates, respectively. The results presented in Figs.~\ref{Fig: Velocity profile}(a--d) feature a mean velocity deficit near $x=0$, which is due to the wake of the spray injector, similar to the results presented in \cite{petry2022quantification}. Also, the mean velocity profiles are nearly symmetric for the test conditions without a perforated plate; however, these profiles are nearly asymmetric for test conditions with one and two perforated plates, which are similar to those reported in \cite{kheirkhah2015consumption}.

The results in Figs.~\ref{Fig: Velocity profile}(e--h) show that, for all tested mean bulk flow velocities and for the majority of the horizontal locations, the RMS of the streamwise velocity for one perforated plate is smaller than that for two perforated plates, which agrees with the results of past investigations, see for example \cite{kheirkhah2015consumption}. The results in Figs.~\ref{Fig: Velocity profile}(e--h) also show that the values of $u^\prime$ for no perforated plate are larger than those of one perforated plate and close to those of two perforated plates. It is speculated that the reason for the values of $u^\prime$ for no perforated plate being relatively large is due to the turbulence generated by the supporting bars of the tube-holder shown in Fig.~\ref{Fig:Nozzle geometry}(a) and the wake of the spray injector. In order to assess this speculation, HWA experiments for a free jet (without the spray injector and without the tube-holder) were performed (not presented as test conditions in Table~\ref{tab:tested onditions}), and the values of $u^\prime$ were obtained. These were significantly smaller than those for the first turbulence generation mechanism. For example, for $U=7.0$~m/s and at $x =0$, $u^\prime=0.34$~m/s for a free jet without the spray injector and the tube-holder; however, this parameter is $1.07$~m/s with the injector and tube-holder installed (i.e. the first turbulence generation mechanism). We speculate the reason for the values of $u^\prime$ being relatively small for the second turbulence generation mechanism (i.e. one perforated plate) than those for the first turbulence generation mechanism (zero perforated plate) is the break up of the eddies generated in the wake of the tube-holder bars by the perforated plate. The decrease of the RMS velocity fluctuations by addition of perforated plates has been reported in for example \cite{wang2019development}

\begin{figure}
	\centerline{\includegraphics[width=1\textwidth]{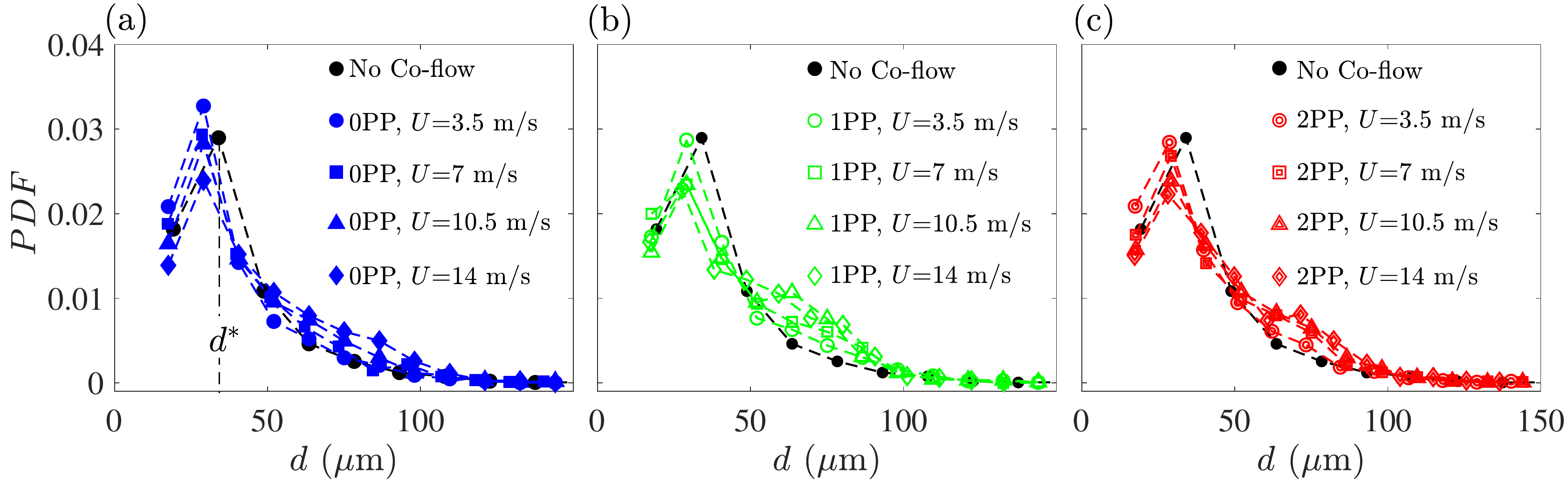}}
	\caption{(a--c) are the probability density functions of the droplet diameter for no perforated plate, one perforated plate, and two perforated plates, respectively. The black circular data points are the PDF of the no co-flow test condition and is repeating in (a--c) for comparison purposes.}
	\label{Fig: droplet size}
\end{figure}

\begin{figure}
	\centerline{\includegraphics[width=0.95\textwidth]{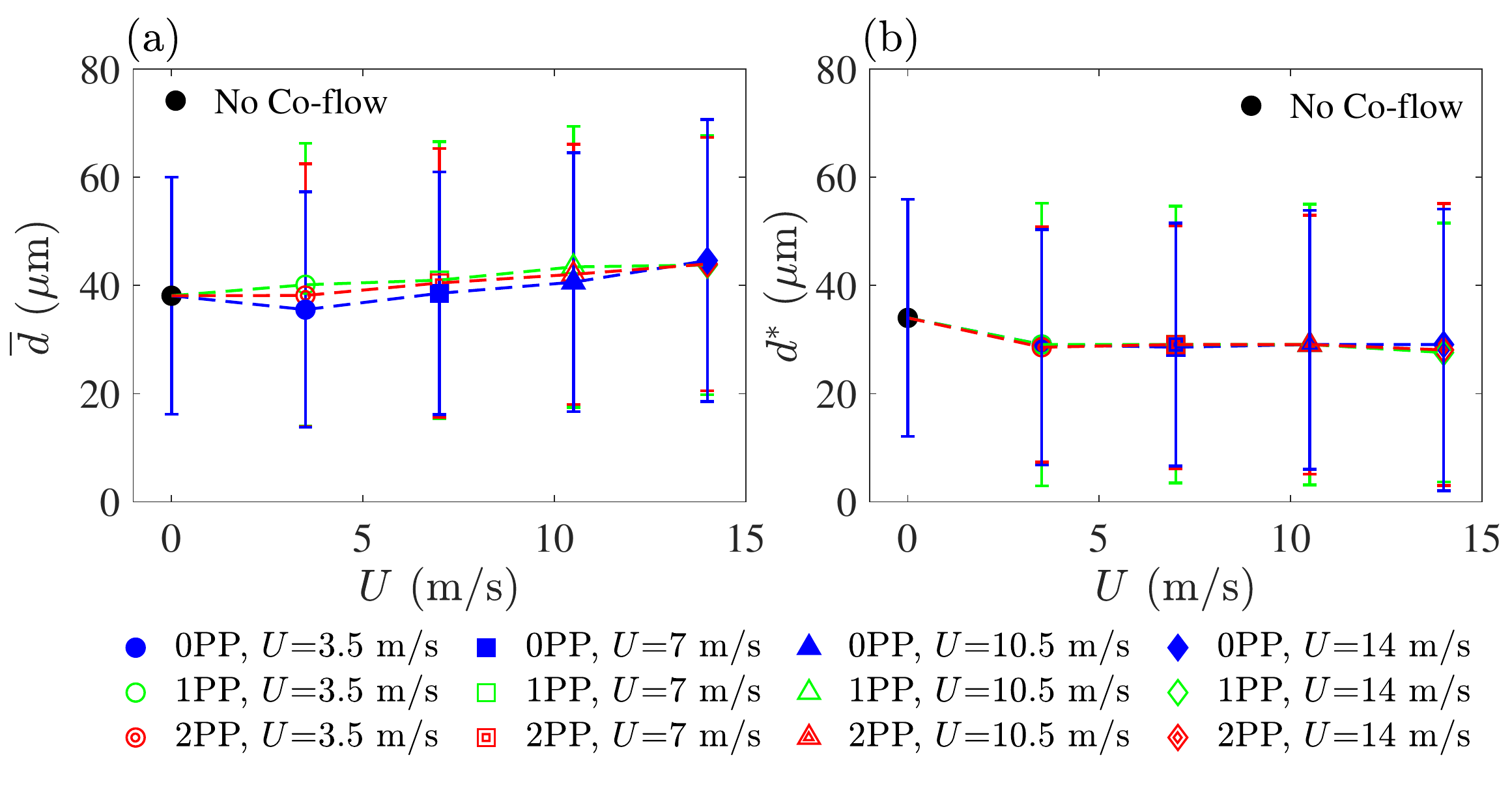}}
	\caption{(a) and (b) are the variations of the mean and most probable droplet diameter versus the mean bulk flow velocity for all test conditions.}
	\label{Fig: Droplet mean diameter}
\end{figure}

The PDF of the droplet diameter for the first, second, and third turbulence generation mechanisms are presented in Figs.~\ref{Fig: droplet size}(a--c), respectively. For comparison purposes, the PDF of the droplet diameter for the no co-flow test condition is overlaid on Fig.~\ref{Fig: droplet size}(a--c) using the black circular data symbol. For the probability density function calculations, an 11.5~$\mu$m droplet diameter bin size was used, since this led to the best presentation of the results. For all test conditions, the mean ($\overline{d}$) and most probable ($d^*$) droplet diameters were obtained and presented in Figs.~\ref{Fig: Droplet mean diameter}(a)~and~(b), respectively. In these figures, the error bars present twice the RMS of the droplet diameter fluctuations. The results presented in Fig.~\ref{Fig: droplet size} show that the presence of the co-flow and increasing the mean bulk flow velocity do not significantly change the PDF of the droplet diameter. Also, comparison of the results presented in Fig.~\ref{Fig: droplet size} suggests that changing the turbulence generation mechanism does not change the PDF of the droplet diameter noticeably. Although the droplets diameter PDF estimated in the entire domain of investigation is not sensitive to the tested mean bulk flow velocity and the utilized turbulence generation mechanism, it is yet to be investigated how/if these parameters influence the PDF of the droplet diameter within the clusters and voids, which are studied in the following subsections.


\subsection{The droplets degree of clustering}\label{subsec:Resultsdegreeofclustering}

\begin{figure}
	\centerline{\includegraphics[width=1\textwidth]{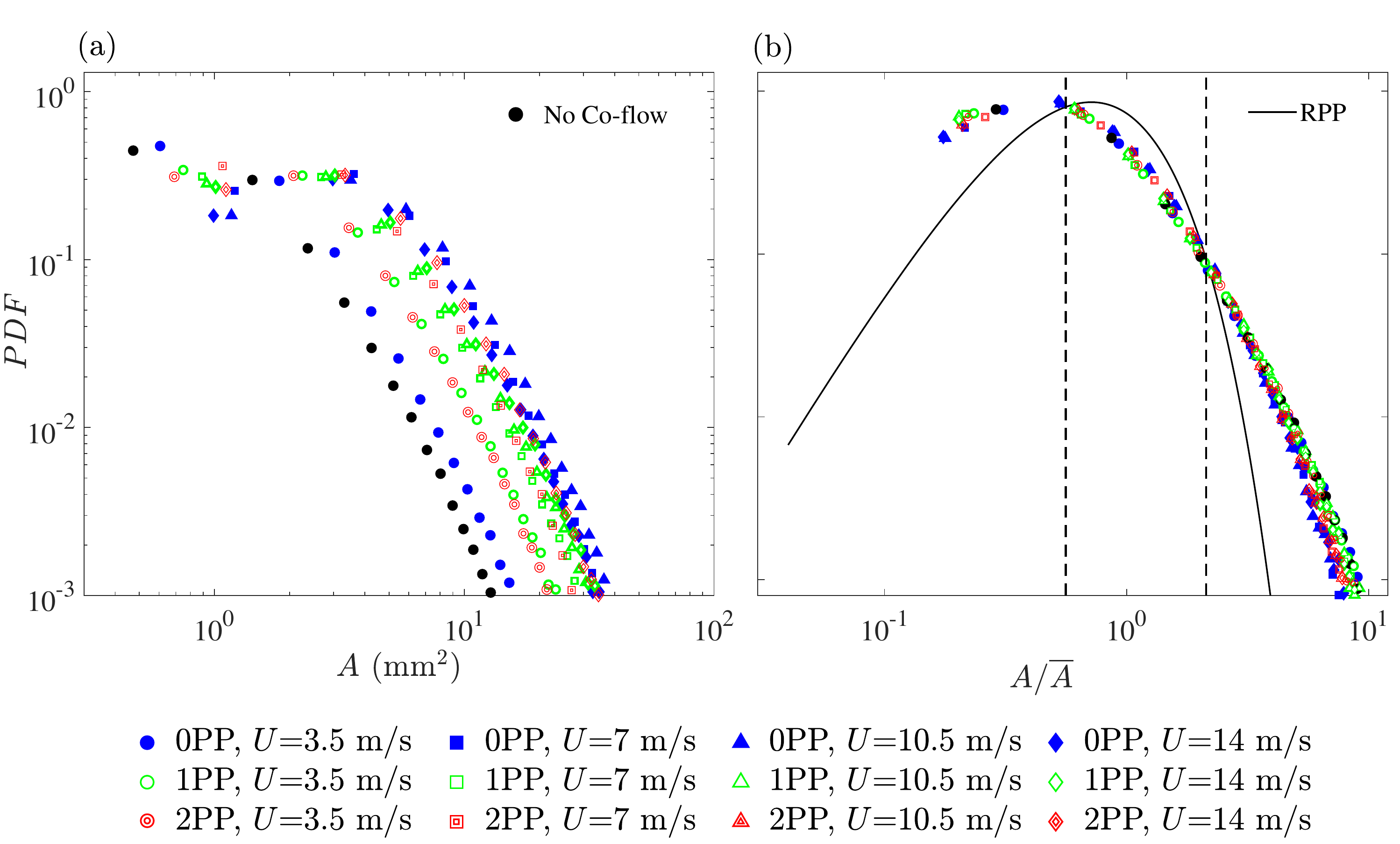}}
	\caption{(a) The probability density functions of the Vorono\"{i} cells area, $PDF(A)$, for all test conditions. (b) is the PDF of Vorono\"{i} cells area normalized by the mean area. The solid black curve in (b) is the PDF of the normalized Vorono\"{i} cells area provided they were distributed following a Random Poison Process, with the formulation given in Eq.~(\ref{eq:random}). The dashed lines in (b) correspond to $A/\overline{A} = 0.5$ and 2.2.}
	\label{Fig: Voronoi cell}
\end{figure}

The Vorono\"{i} cells were used to study the droplets degree of clustering, following the procedure discussed in subsection~\ref{sec:datareduction}. For all test conditions, the PDFs of the Vorono\"{i} cells area ($A$) are calculated and presented in Fig.~\ref{Fig: Voronoi cell}(a). Also, the PDFs of the Vorono\"{i} cells area normalized by the mean area ($A/\overline{A}$) are presented in Fig.~\ref{Fig: Voronoi cell}(b) for all test conditions. As can be seen, the PDFs of $A/\overline{A}$ collapse for all tested conditions, which is similar to the results presented in \cite{obligado2014preferential} and \cite{monchaux2010preferential}. Overlaid on Fig.~\ref{Fig: Voronoi cell} is the PDF of the normalized area of the Vorono\"{i} cells provided these cells are spatially distributed following the Random Poison Process, with the formulation of $PDF_\mathrm{RPP}$ presented in Eq.~(\ref{eq:random}). The results in Fig.~\ref{Fig: Voronoi cell} show that, for all test conditions, the PDFs of $A/\overline{A}$ intersect with $PDF_\mathrm{RPP}$ at $A/\overline{A} = 0.5$ and 2.2, which are shown by the vertical dashed lines in Fig.~\ref{Fig: Voronoi cell}(b) and are similar to those shown in Fig.~\ref{Fig: Data Reduction}(c). Using the above normalized areas and following the procedure presented in subsection~\ref{subsec:Miedatareduction}, the clusters and voids were identified, and their degree of clustering is studied below.

\begin{figure}
	\centerline{\includegraphics[width=1\textwidth]{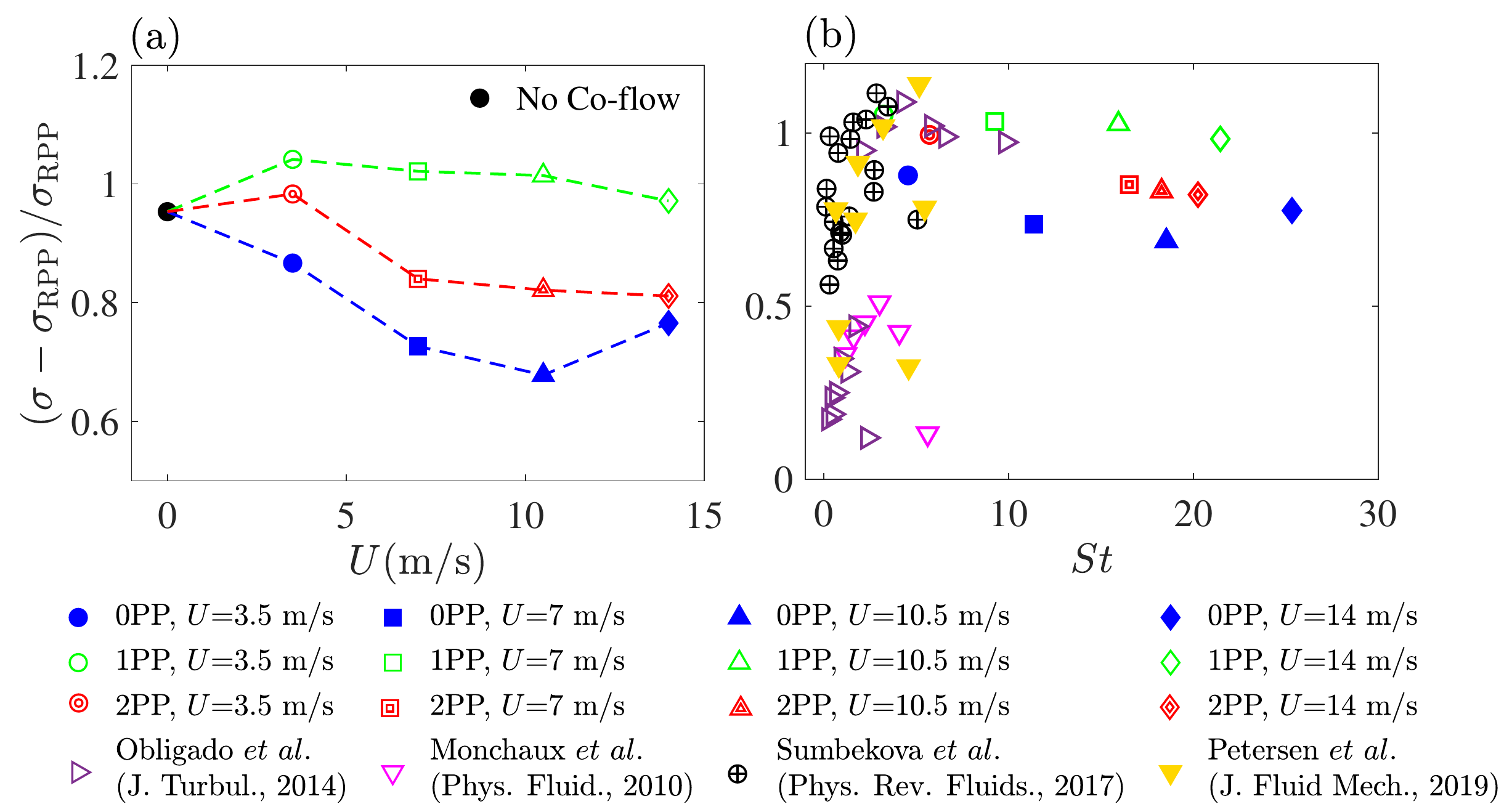}}
	\caption{(a) and (b) are the degree of clustering versus the mean bulk flow velocity and the Stokes number, respectively. Overlaid on (b) are the results of \cite{obligado2014preferential,monchaux2010preferential,sumbekova2017preferential,petersen2019experimental}.}
	\label{Fig:clustering}
\end{figure}

The degree of clustering, $(\sigma-\sigma_\mathrm{RPP})/\sigma_\mathrm{RPP}$, is presented in Fig.~\ref{Fig:clustering}(a) for all test conditions. The results in the figure show that, for the mean bulk flow velocities larger than or equal to 3.5~m/s, the presence of the turbulent co-flow increases the degree of clustering for one perforated plate (compared to no co-flow); however; for test conditions with zero and two perforated plates, the presence of the co-flow generally decreases the degree of clustering. For comparison purposes, the variation of the degree of clustering versus the Stokes number is presented in Fig.~\ref{Fig:clustering}(b). In addition to the results of the present study, those of \cite{obligado2014preferential,monchaux2010preferential,sumbekova2017preferential,petersen2019experimental} are also overlaid on the figure. The results show that increasing the Stokes number from 0 to about 10 increases the degree of clustering from about 0 to 1.2; however, this parameter nearly plateaus at around 0.7 to 1 for $St \gtrsim 5$. \cite{wang2020reynolds} performed 3D Direct Numerical Simulation (DNS) of particles clustering in homogeneous and isotropic turbulent flow in a box for Stokes numbers ranging from about 0 to 7. They estimated the standard deviation of the 3D Vorono\"{i} cells volume. Results of \cite{wang2020reynolds} show that the standard deviation of the 3D Vorono\"{i} cells volume increases by increasing the Stokes number from about 0 to 1. However, further increase of the Stokes number to about 7 plateaus the standard deviation of the Vorono\"{i} cells volume. Acknowledging that the analysis of \cite{wang2020reynolds} is performed for the RMS of the Vorono\"{i} cells volume (which is different from the degree of clustering that pertains to the Vorono\"{i} cells area), the reported trend in \cite{wang2020reynolds} follows that presented in Fig.~\ref{Fig:clustering}(b) for matching Stokes numbers. Nevertheless, to our best knowledge, the plateau of the degree of clustering with increasing the Stokes number for $St \gtrsim 7$ is reported in the present study for the first time.

\subsection{The geometrical characteristics of clusters and voids}\label{subsec:Resultsclustersandvoidsgeometries}

The spatial distribution and length scale of the clusters and voids as well as how these are influenced by the non-dimensional parameters are discussed in the following subsections.

\subsubsection{Characteristics of clusters and voids spatial distribution}

Although the analysis presented in subsection~\ref{subsec:Resultsdegreeofclustering} allows for understanding the droplets degree of clustering, it does not provide insight into the characteristics of the clusters and voids themselves. In this subsection, it is of interest to investigate if the clustering of the clusters and voids occurs, how the clusters and voids are spatially distributed, what the length scale of clusters and voids are,  and how/if the non-dimensional parameters influence the above.

\begin{figure}	
	\centerline{\includegraphics[width=0.7\textwidth]{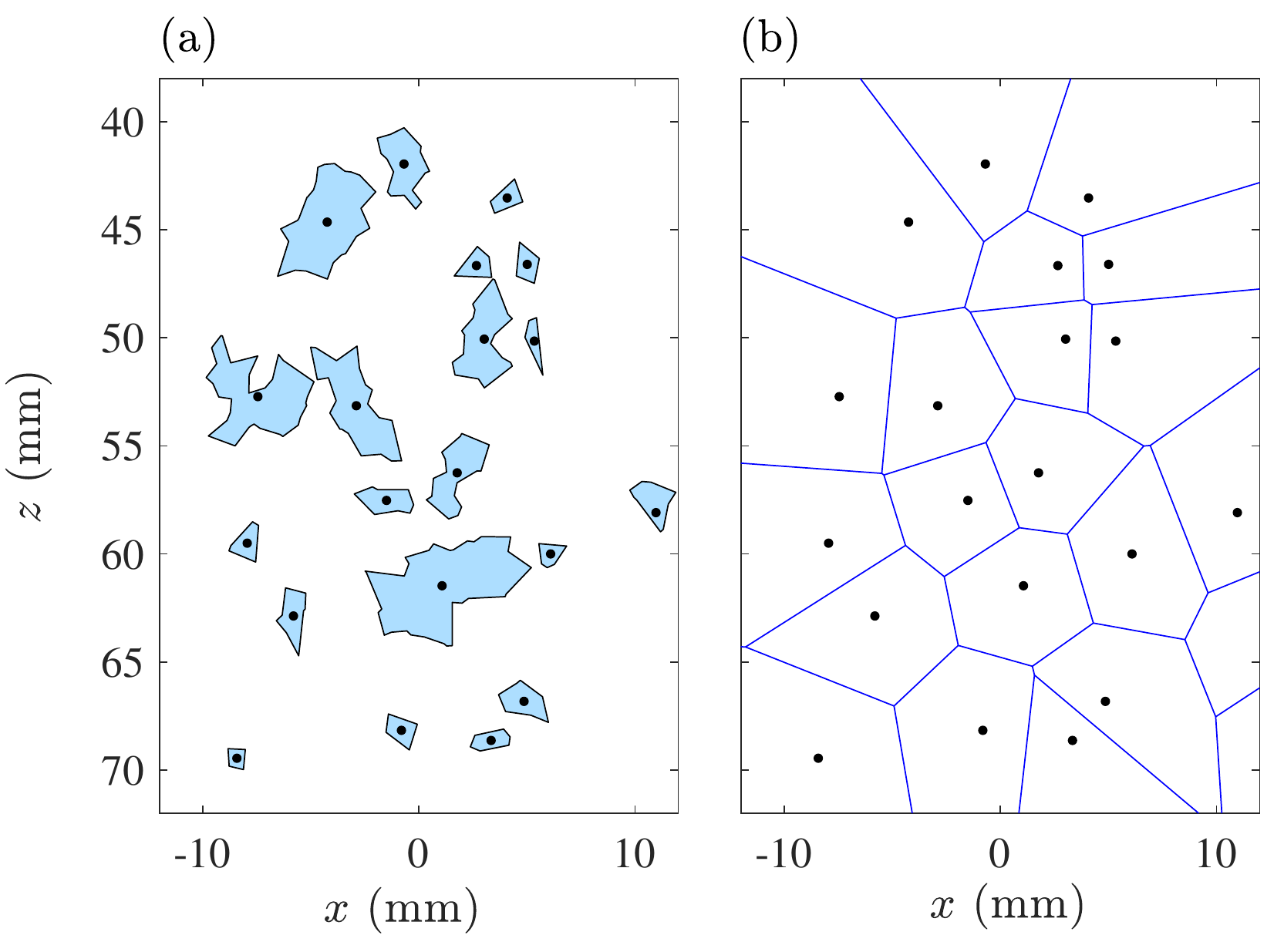}}
	\caption{(a) Representative clusters. (b) The Vorono\"{i} cells formed by the clusters center of area. In both (a) and (b), the clusters center of area are shown by the solid black data symbol. The results in (a) and (b) correspond to the no co-flow test condition.}
	\label{Fig: 2nd Voronoi analysis}
\end{figure}

\begin{figure}
\centerline{\includegraphics[width=0.7\textwidth]{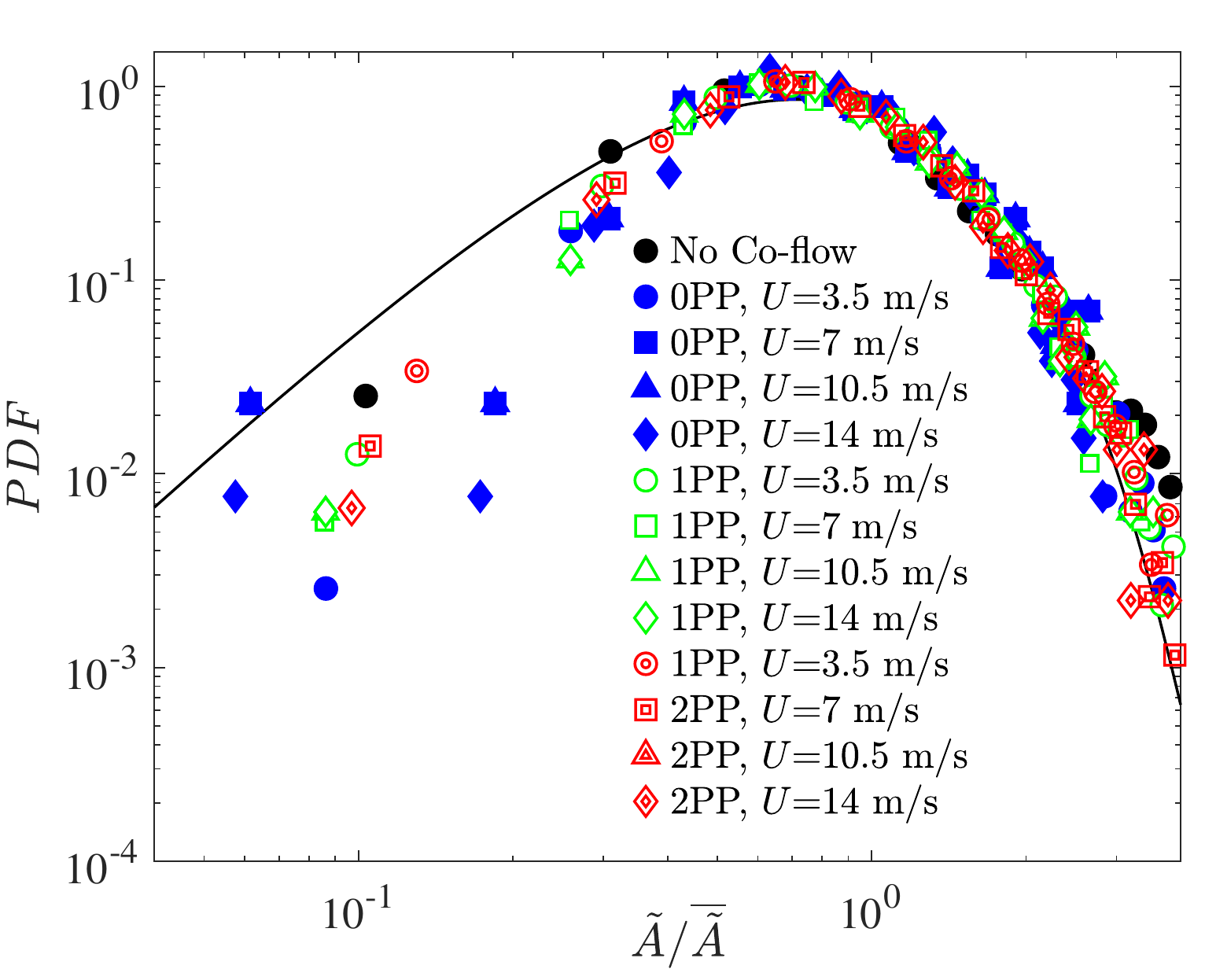}}
	\caption{The probability density functions of the normalized Vorono\"{i} cells area formed by the clusters center of area. The black solid curve is the $PDF_\mathrm{RPP}$.}
	\label{Fig: PDF of 2nd voronoi}
\end{figure}

A representative spatial distribution of the clusters for the no co-flow test condition is presented in Fig.~\ref{Fig: 2nd Voronoi analysis}(a), see the blue regions in the figure. The centers of the area of the clusters were obtained and are shown by the black circular symbols in both Figs.~\ref{Fig: 2nd Voronoi analysis}(a and b). The Vorono\"{i} cells associated with the center of areas of the clusters were obtained and are shown in Fig.~\ref{Fig: 2nd Voronoi analysis}(b) using the blue lines. For all test conditions, the Probability Density Function of the Vorono\"{i} cells area obtained using the center area of the clusters ($\tilde{A}$) divided by its mean value ($\overline{\tilde{A}}$) is presented in Fig.~\ref{Fig: PDF of 2nd voronoi}. Also overlaid on the figure is the $PDF_\mathrm{RPP}$. As can be seen, the PDFs of $\tilde{A}/\overline{\tilde{A}}$ for all test conditions nearly collapse on $PDF_\mathrm{RPP}$, suggesting that the centers of area of the clusters are distributed following a Random Poison Process. Thus, the clustering of the clusters does not occur for the conditions tested in the present study. A similar analysis was performed to understand if the clustering of the voids occurs. Since the voids are positioned at the periphery of the domain of investigation, see for example Fig.~\ref{Fig: Data Reduction}(f), the vertices of the Vorono\"{i} cells generated from the voids center of area cannot be accessed. Thus, for the spray and co-flow configuration of the present study, the clustering of the voids cannot be investigated. 

\begin{figure}
	\centerline{\includegraphics[width=1\textwidth]{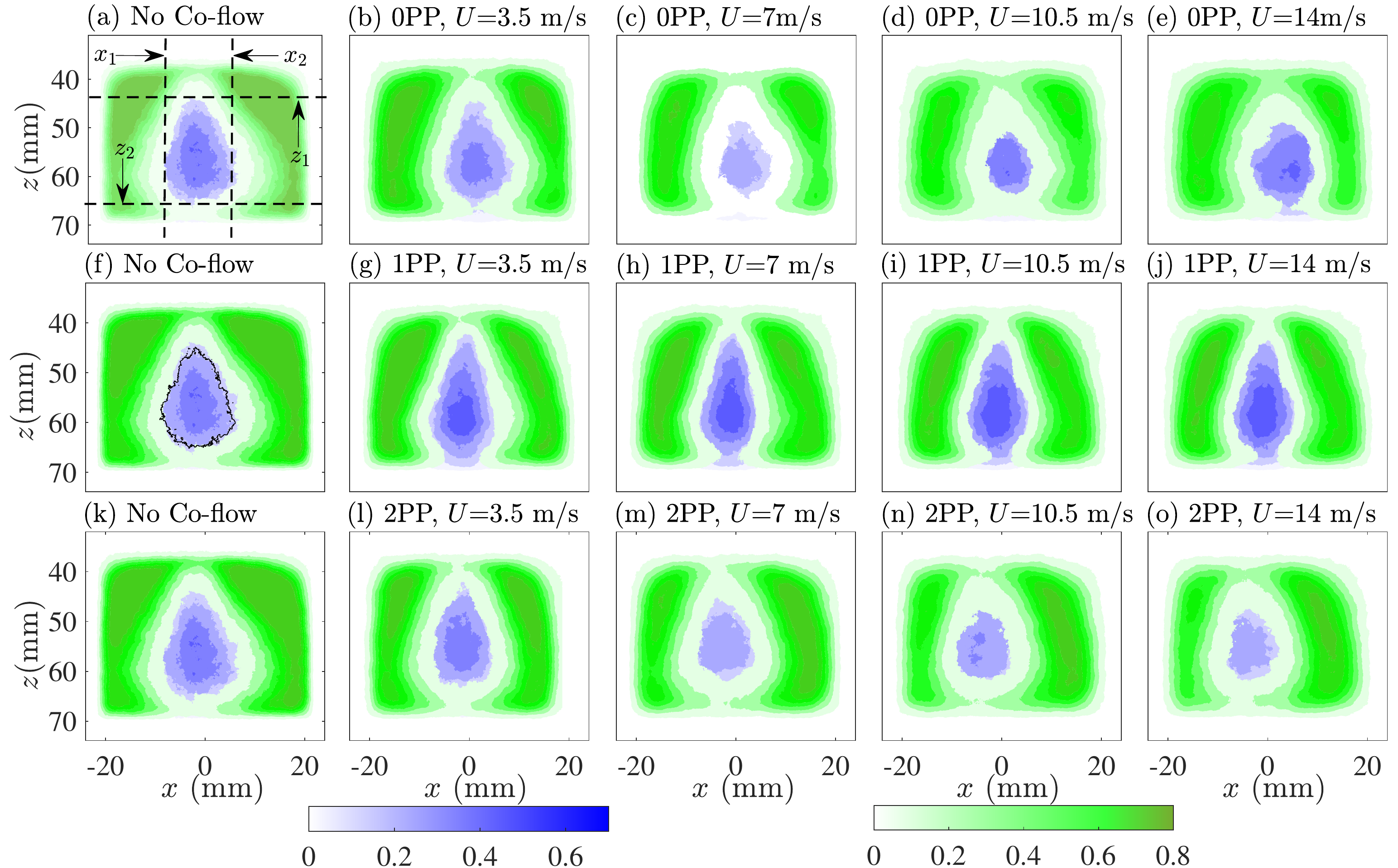}}
	\caption{The blue and green contours present the probabilities of finding clusters and voids, respectively. The first to fifth columns correspond to no co-flow, $U= 3.5$, 7.0, 10.5, and 14.0~m/s, respectively. The results in (b--e), (g--j), and (l--o) pertain to the first to third turbulence generation mechanisms, respectively.}
	\label{Fig: cluster-void location}
\end{figure}

\begin{figure}
	\centerline{\includegraphics[width=0.8\textwidth]{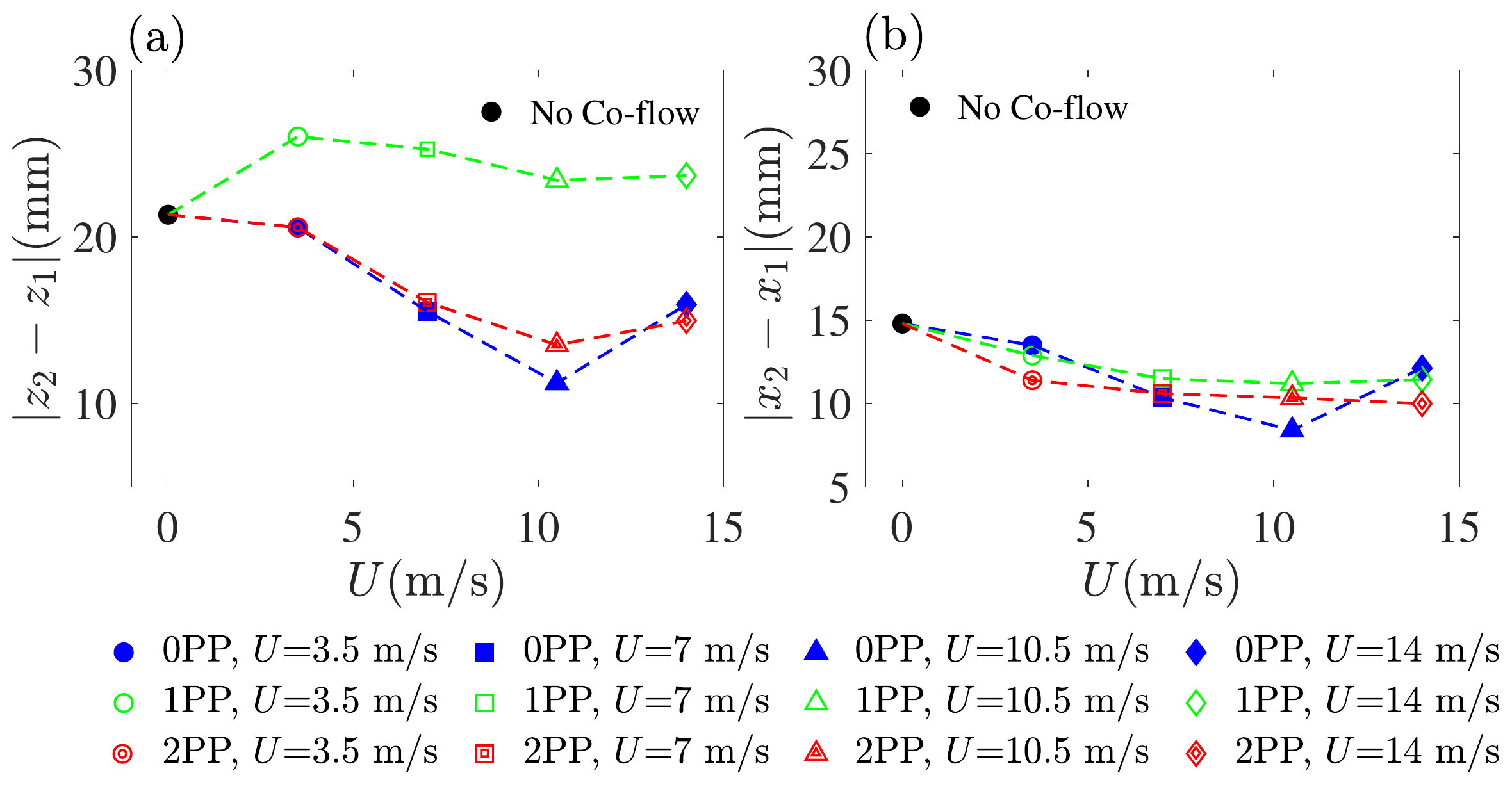}}
	\caption{(a) and (b) are the variations of the vertical and horizontal extents of the regions that feature more than 20\% probability of finding clusters versus the mean bulk flow velocity, respectively.}
	\label{Fig: cluster location versus U}
\end{figure}

The spatial probabilities of finding clusters and voids were calculated using their geometrical locations, such as those shown in Fig.~\ref{Fig: Data Reduction}(e~and~f). For clusters, the blue regions, for example those in Fig.~\ref{Fig: Data Reduction}(e), were assigned a unity value and the rest of the domain of investigation was assigned zero. Then, all binarized results were averaged and was referred to as the spatial probability of finding clusters. A similar procedure was followed to obtain the probability of finding voids in the domain of investigation using the location of voids such as that shown in Fig.~\ref{Fig: Data Reduction}(f). The spatial probabilities of finding clusters and voids were overlaid together and presented in Fig.~\ref{Fig: cluster-void location} using the blue and green color contours, respectively. In Fig.~\ref{Fig: cluster-void location}, the first column presents the results for the no co-flow test condition, is repeating, and is presented for comparison purposes. The second to fifth columns correspond to the mean bulk flow velocities of 3.5, 7.0, 10.5, and 14.0~$\mathrm{m/s}$, respectively. The results in Figs.~\ref{Fig: cluster-void location}(b--e), (g--j), and (l--o) correspond to test conditions with zero, one, and two perforated plates, respectively. As can be seen, the probability of finding clusters is maximized at the spray core, which is similar to the findings of \cite{zimmer2003simultaneous} and \cite{jedelsky2018air}. Compared to the clusters, the probability of finding voids is maximized at the periphery of the domain of investigation. It can also be seen that the shape of the region with a relatively large probability of finding the clusters change by changing the test condition. In order to quantify this, the contour that corresponds to 20\% probability of finding the clusters was obtained for all test condition, with that for the no co-flow test condition overlaid by the black solid curve in Fig.~\ref{Fig: cluster-void location}(f). Sample boundaries of such contours for the no co-flow test condition are shown by $x_1$, $x_2$, $z_1$, and $z_2$ in Fig.~\ref{Fig: cluster-void location}(a). The horizontal ($x_2-x_1$) and vertical ($z_2-z_1$) extents of the above boundaries were obtained for all test conditions, with the results shown in Fig.~\ref{Fig: cluster location versus U}(a)~and~(b), respectively. As can be seen in Fig.~\ref{Fig: cluster location versus U}(b), the presence of the co-flow (compared to the no co-flow test condition) decreases the horizontal extents of the regions with a large probability (more than 20\%) of finding clusters. A similar observation is made for the vertical extent of these regions for the first and third turbulence generation mechanisms. However, the presence of the co-flow elongates the above regions vertically for the second turbulence generation mechanism.

\begin{figure}
	\centering
	\includegraphics[width=0.7\textwidth]{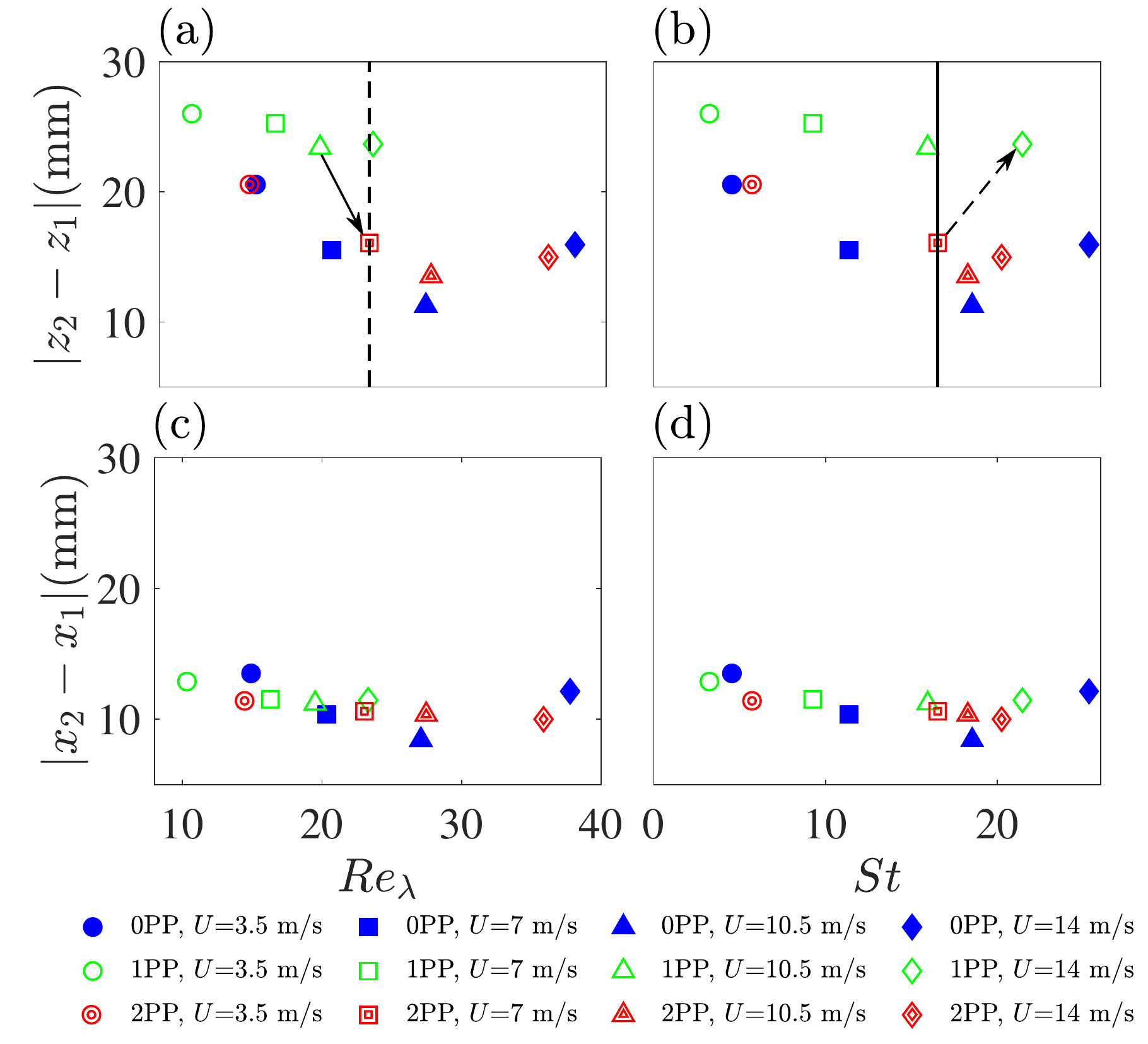}
	\caption{(a) and (b) are the vertical extent of the regions that feature more than 20\% probability of finding clusters versus $Re_\lambda$ and $St$, respectively. (c) and (d) are the horizontal extent of the regions that feature more than 20\% probability of finding clusters versus $Re_\lambda$ and $St$, respectively.}
	\label{Fig: cluster location vs st}
\end{figure}

In order to study the reason for the above observations, $z_2-z_1$ versus $Re_\lambda$ and $St$ are presented in Figs.~\ref{Fig: cluster location vs st}(a)~and~(b), respectively. Also, variations of $x_2-x_1$ versus $Re_\lambda$ and $St$ are presented in Figs.~\ref{Fig: cluster location vs st}(c)~and~(d), respectively. Comparison of the results in Fig.~\ref{Fig: cluster location vs st}(a and b) with those in Fig.~\ref{Fig: cluster location vs st}(c and d) suggests, although $z_2-z_1$ is significantly influenced by $Re_\lambda$ and $St$, $x_2-x_1$ does not change remarkably with these non-dimensional parameters. For a fixed value of $Re_\lambda$, see the dashed line in Fig.~\ref{Fig: cluster location vs st}(a), increasing the Stokes number, see the corresponding dashed arrow in Fig.~\ref{Fig: cluster location vs st}(b), increases $z_2-z_1$. However, for a fixed value of the Stokes number, see the solid line in Fig.~\ref{Fig: cluster location vs st}(b), increasing the Reynolds number decreases $z_2-z_1$, see the solid arrow in Fig.~\ref{Fig: cluster location vs st}(a). This means that, for test conditions that feature similar background $Re_\lambda$, increasing the inertia of the droplets increases the probability of finding clusters at a relatively large range of vertical distances from the injector. Although, for a fixed Reynolds number, it is anticipated that increasing the Stokes number should increase the probability of finding droplets at large distances from the injector, the results in Fig.~\ref{Fig: cluster location vs st}(b) suggest that the droplets continue to cluster at large distances from the injector.

\begin{figure}
	\centerline{\includegraphics[width=0.85\textwidth]{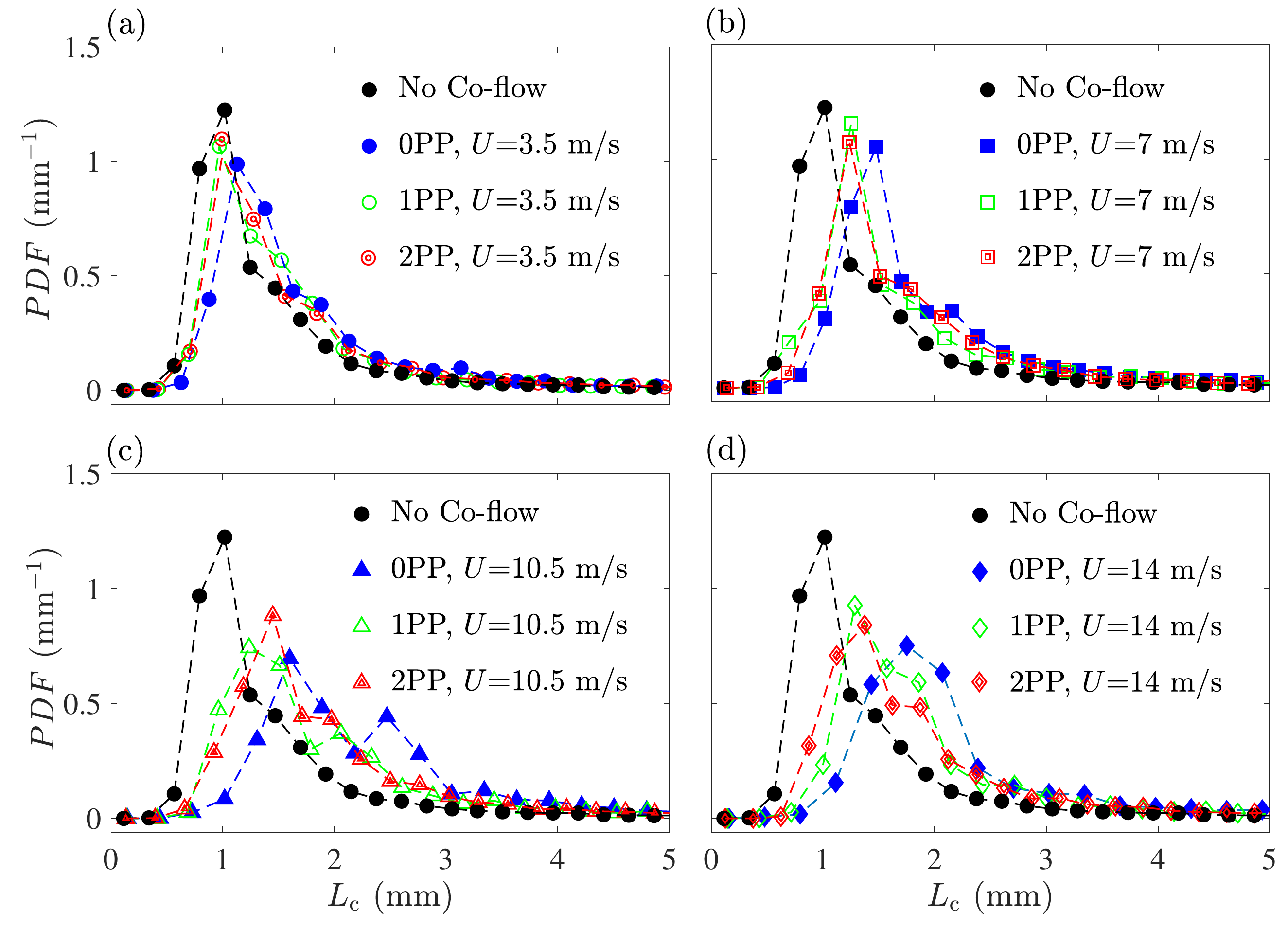}}
	\caption{(a--d) are the PDFs of the cluster length scale for mean bulk flow velocities of $U=3.5$, 7.0, 10.5, and 14.0~m/s, respectively. Overlaid by the black circular data symbol in (a--d) is the PDF of the cluster length scale for the no co-flow test condition.}
	\label{Fig: PDF length scale of clusters}
\end{figure}

\subsubsection{Length scale of clusters and voids}
\begin{figure}
	\centerline{\includegraphics[width=0.85\textwidth]{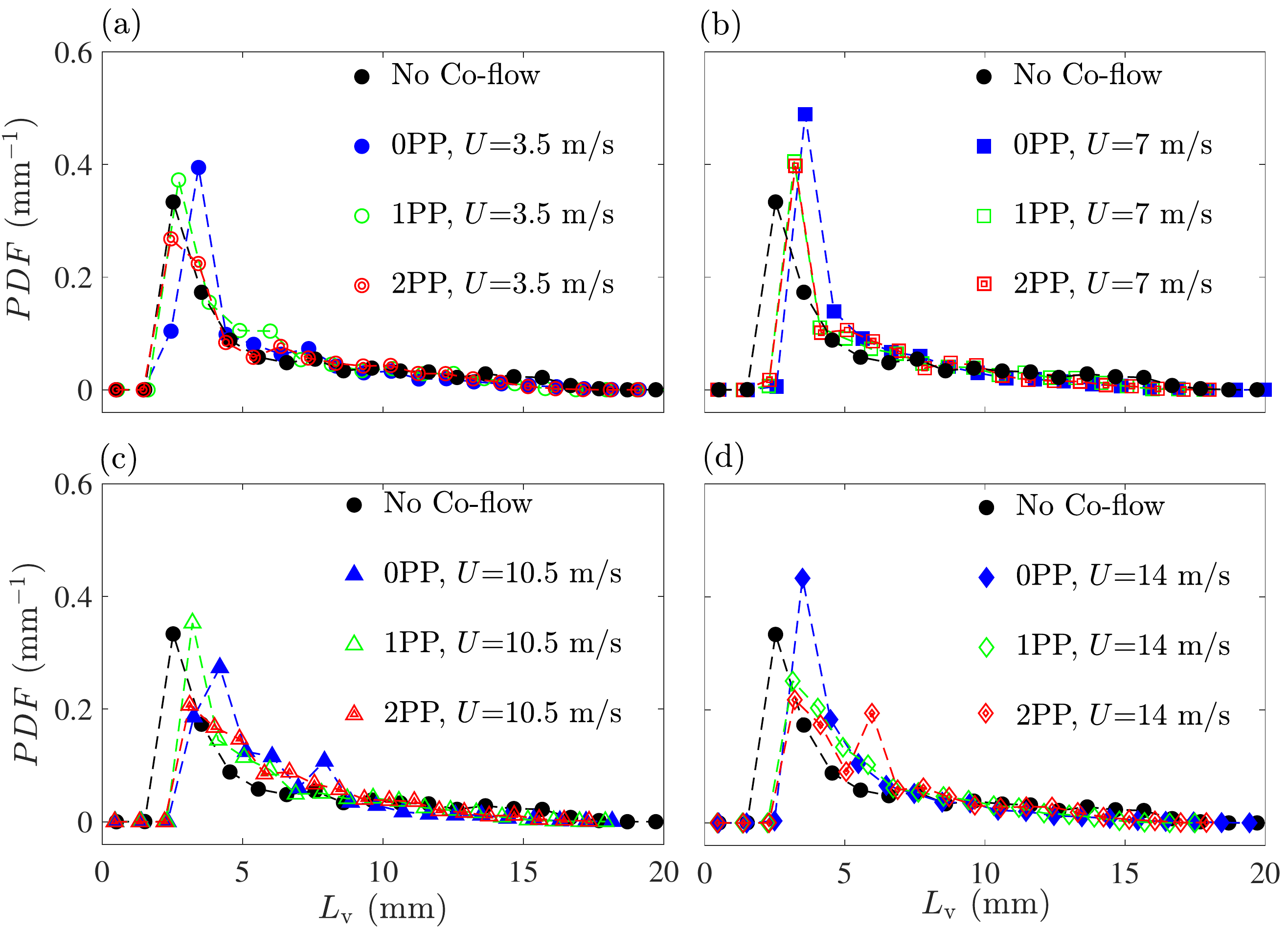}}
	\caption{(a--d) are the PDFs of the void length scale for the mean bulk flow velocities of 3.5, 7.0, 10.5, and 14.0~m/s, respectively. The black circular data symbol is the PDF of the void length scale for the no co-flow test condition.}
	\label{Fig: PDF void length scale}
\end{figure}

Following \cite{petersen2019experimental}, the cluster and void length scales were defined as $L_\mathrm{c} = \sqrt{A_\mathrm{c}}$ and $L_\mathrm{v} = \sqrt{A_\mathrm{v}}$, respectively, with $A_\mathrm{c}$ and $A_\mathrm{v}$ being the cluster and void areas. The PDFs of $L_\mathrm{c}$ and $L_\mathrm{v}$ are presented in Figs.~\ref{Fig: PDF length scale of clusters}~and~\ref{Fig: PDF void length scale}, respectively. The results in Figs.~\ref{Fig: PDF length scale of clusters}(a--d)~and~\ref{Fig: PDF void length scale}(a--d) correspond to mean bulk flow velocities of 3.5, 7.0, 10.5, and 14.0~m/s, respectively. For comparison purposes, the PDFs of the cluster and void length scales corresponding to the no co-flow condition is presented by the solid black circular data symbol on the figures. The results in Fig.~\ref{Fig: PDF length scale of clusters} show that, for a given turbulence generation mechanism, the most probable cluster and void length scales increase by increasing the mean bulk flow velocity. It is also observed that, for a given mean bulk flow velocity, the most probable cluster and void length scales are largest for the test conditions with no perforated plates. Compared to the effects of $U$ and turbulence generation mechanism on the most probable cluster length scale, the most probable void length scale does not change by changing the above parameters.

\begin{figure}
	\centerline{\includegraphics[width=0.8\textwidth]{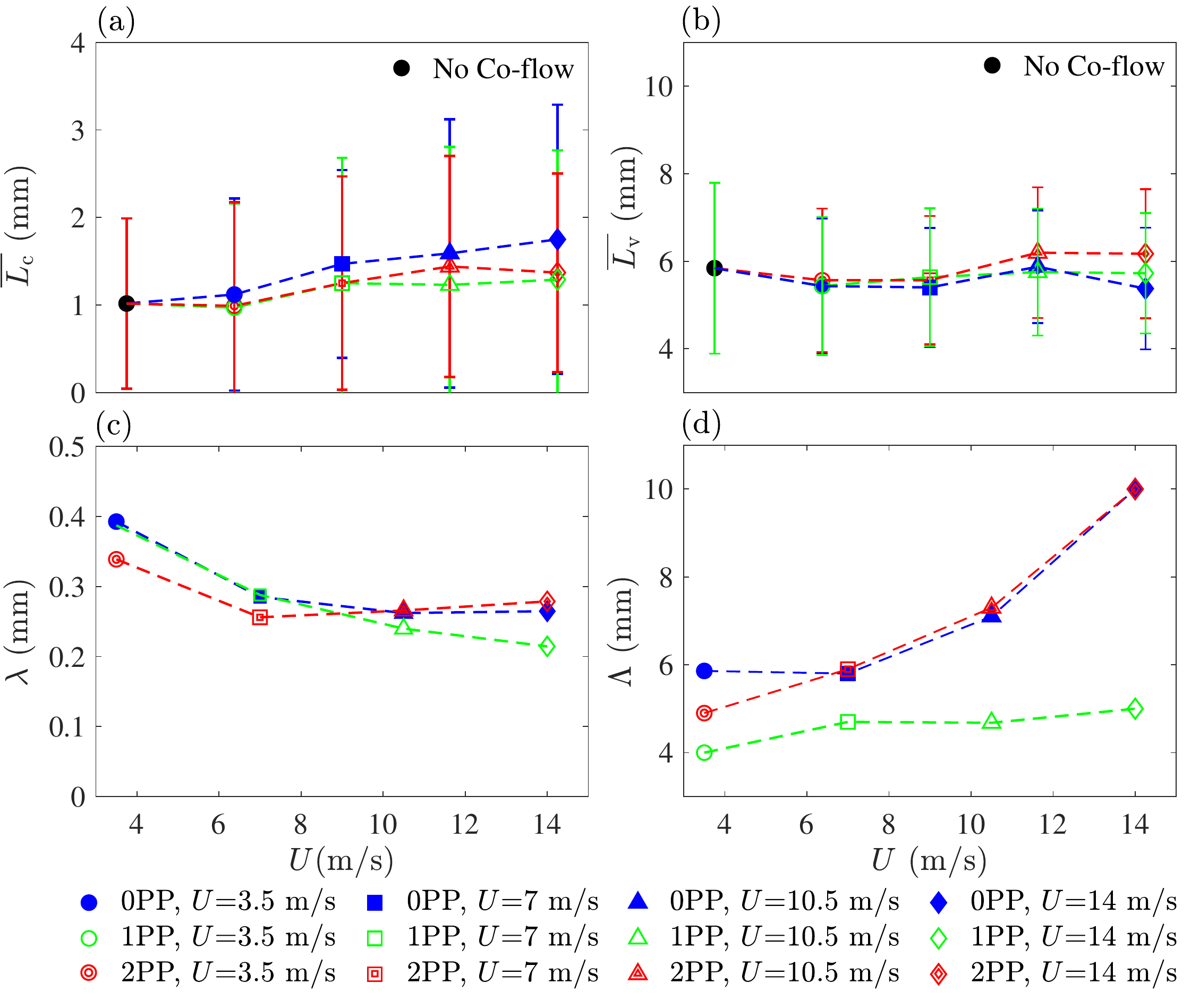}}
	\caption{(a) and (b) are the mean cluster and void length scales versus the mean bulk flow velocity, respectively. (c) and (d) are the Taylor and integral length scales versus the mean bulk flow velocity, respectively.}
	\label{Fig: mean clusters and voids length}
\end{figure}

Although the above analysis related to the most probable cluster and void length scales are of importance, majority of past investigations studied the relations between the mean values of the above length scales and the non-dimensional parameters, such as Taylor length scale based-Reynolds and Stokes numbers. The mean cluster and void length scales versus the mean bulk flow velocity are presented in Figs.~\ref{Fig: mean clusters and voids length}(a) and (b), respectively. The error bars in the figures are twice the standard deviation of the corresponding data. For comparison purposes, the Taylor and integral length scales are also presented in Figs.~\ref{Fig: mean clusters and voids length}(c)~and~(d), respectively. As can be seen, increasing the mean bulk flow velocity slightly increases the mean cluster size, however, the mean void size remains nearly constant. The results show that cluster size is larger than the Taylor length scale but smaller than the integral length scale. The mean void length scale is, however, close to the integral length scale.

\begin{figure}
	\centerline{\includegraphics[width=0.9\textwidth]{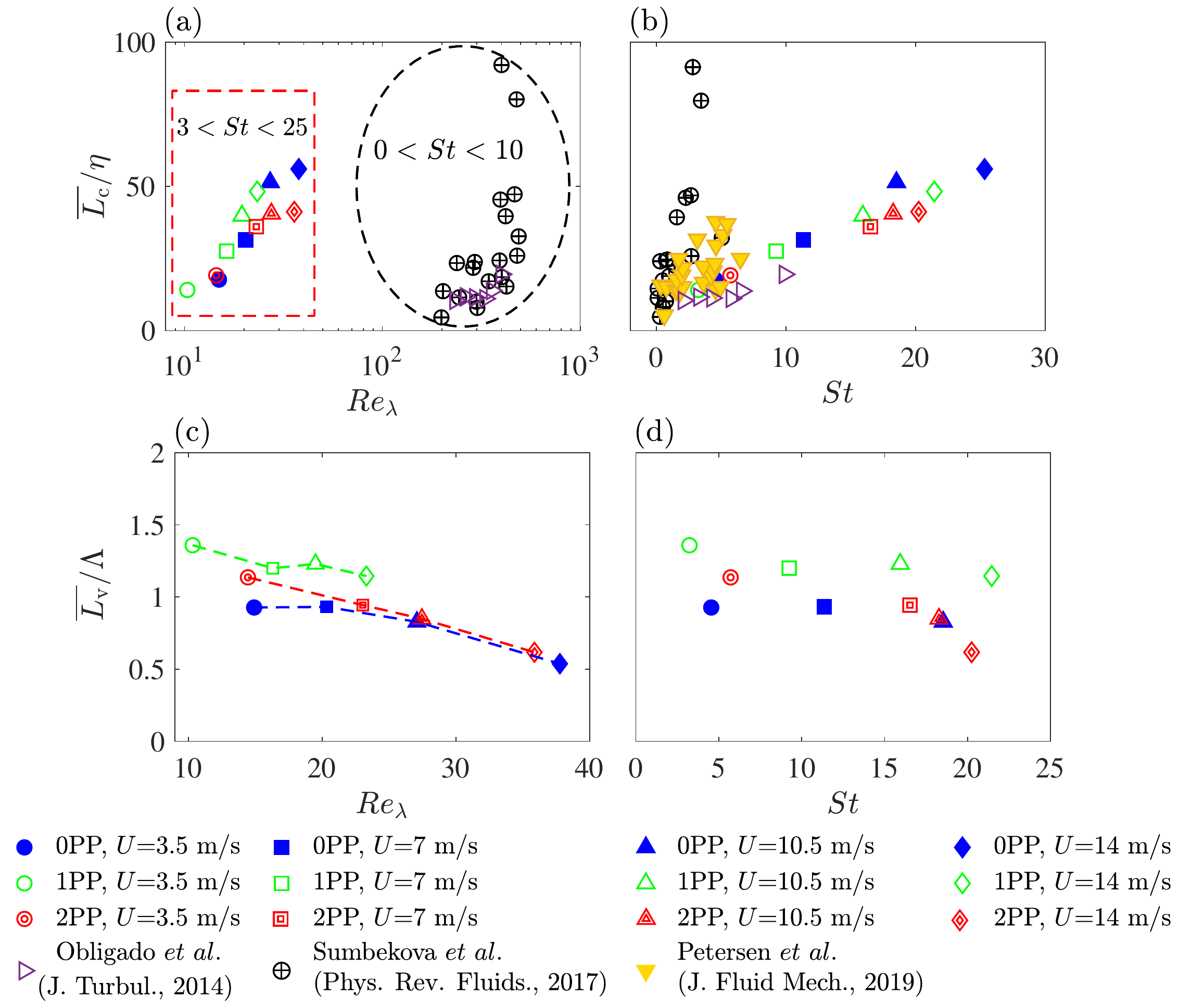}}
	\caption{(a) and (b) are the mean cluster length scale normalized by the Kolmogorov length scale versus $Re_\lambda$. Overlaid on (a and b) are the results of \cite{obligado2014preferential}, \cite{sumbekova2017preferential}, and \cite{petersen2019experimental}. (c and d) are the mean void length scale normalized by the integral length scale versus $Re_\lambda$ and $St$, respectively.}
	\label{Fig: Length scale ratios}
\end{figure}

Variations of $\overline{L_\mathrm{c}}$ normalized by the Kolmogorov length scale versus $Re_\lambda$ is presented in Fig.~\ref{Fig: Length scale ratios}(a). Also, overlaid on the figure are the results of \cite{obligado2014preferential} and \cite{sumbekova2017preferential}, which are highlighted by the black dashed ellipse in the figure and correspond to relatively small Stokes numbers ($0 \lesssim St \lesssim 10$). The results of the present study are highlighted by the dashed red rectangle. As can be seen, the normalized cluster length scale of the present study and those of \cite{obligado2014preferential} and \cite{sumbekova2017preferential} do not follow a trend. Thus, variation of $\overline{L_\mathrm{c}}/\eta$ versus the Stokes number were obtained and presented in Fig.~\ref{Fig: Length scale ratios}(b). The results of \cite{obligado2014preferential,sumbekova2017preferential} as well as \cite{petersen2019experimental} are overlaid on the figure for comparison. As can be seen, the normalized mean cluster length scale follows a trend when presented against the Stokes number. Specifically, it can be seen that increasing $St$ almost linearly increases $\overline{L_\mathrm{c}}/\eta$. Variations of $\overline{L_\mathrm{v}}/\Lambda$ versus $Re_\lambda$ and $St$ are presented in Fig.~\ref{Fig: Length scale ratios}(c)~and~(d), respectively. Compared to the mean cluster length scale that can become about 100 times larger than the Kolmogorov length scale, the results in Fig.~\ref{Fig: Length scale ratios}(c and d) show that the mean void length scale is on the order of the integral length scale. For large Stokes numbers, the droplets feature relatively large inertia, they interact with large scale eddies, and as a result, the droplets position at the periphery of the large scale eddies. This would suggest that, for large Stokes numbers (such as those of the present study), the regions inside the large eddies correspond to voids; and, as a result, these regions size is about the voids length scale, as shown in Fig.~\ref{Fig: Length scale ratios}(d). \cite{yoshimoto2007self} performed DNS of particles interacting with homogeneous and isotropic turbulence in a box, and they, indeed, showed that for $St = 10$, the length scale of voids saturate at the integral length scale of the turbulent flow. The results of the present study extends the DNS of \cite{yoshimoto2007self} from homogeneous and isotropic turbulence in a box to sprays in a turbulent co-flow with Stokes numbers up to 25. 

\subsection{The inter-cluster and inter-void characteristics}\label{subsec:Resultsinternalstructure}

The results discussed in subsections~\ref{subsec:Resultsbackgroundflow}--\ref{subsec:Resultsclustersandvoidsgeometries} allow for understanding the characteristics of the droplets as well as those of clusters and voids individually; however, our understanding related to joint characteristics of the droplets and clusters/voids remains to be developed. In the following, the number density of the droplets inside the clusters and voids as well as the Joint Probability Density Function (JPDF) of the droplets diameter and clusters/voids areas are investigated.

\subsubsection{Number densities of droplets inside clusters and voids}
\label{subsubsec:clusterandvoidsnumberdensities}
\begin{figure}
	\centerline{\includegraphics[width=1\textwidth]{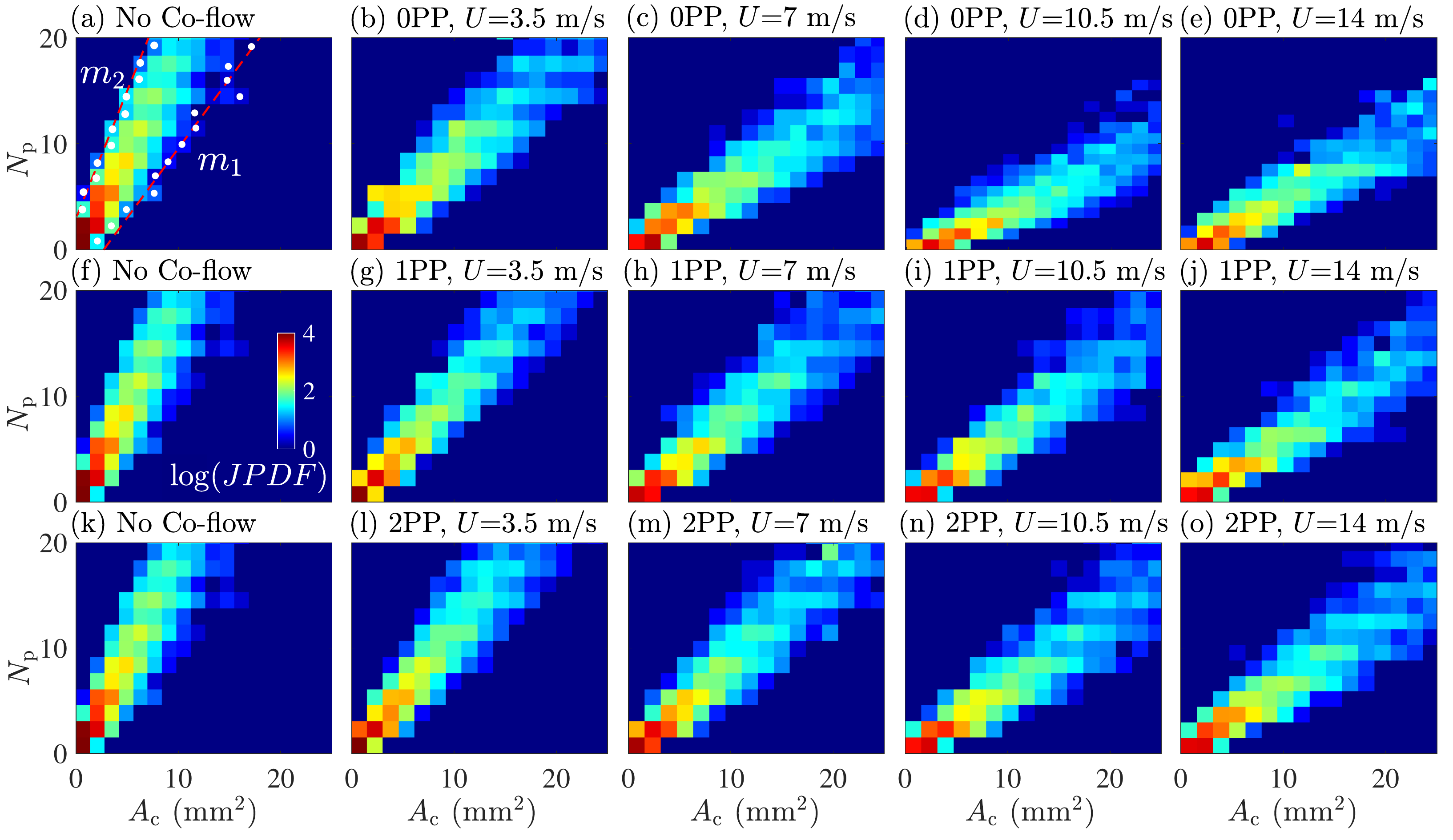}}
	\caption{The logarithmic joint probability density function of the number of droplets within a given cluster area. The first column corresponds to the no co-flow test condition and is repeating in each row for comparison purposes. The second to fifth columns correspond to the mean bulk flow velocities of 3.5, 7.0, 10.5, and 14.0~m/s, respectively. (b--e), (g--j), and (l--o) pertain to the first to third turbulence generation mechanisms, respectively.}
	\label{Fig: Number density contour}
\end{figure}

\begin{figure}
	\centerline{\includegraphics[width=1\textwidth]{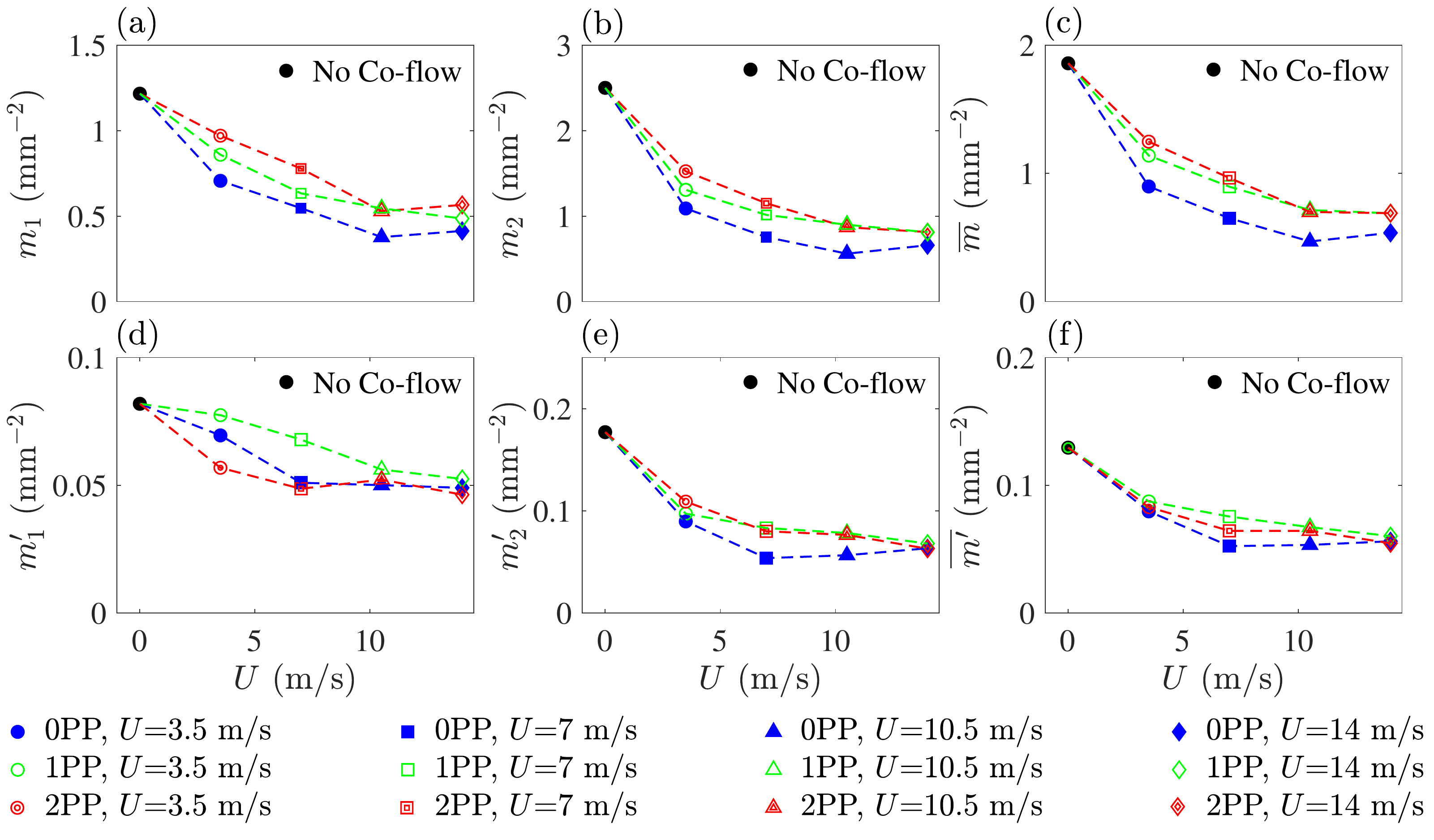}}
	\caption{(a--c) are variations of $m_1$, $m_2$, and $0.5(m_1+m_2)$ versus $U$, respectively. (d--f) are variations of $m^\prime_1$, $m^\prime_2$, and $0.5(m^\prime_1+m^\prime_2)$ versus $U$, respectively.}
	\label{Fig: number density vs u}	
\end{figure}

\begin{figure}
	\centerline{\includegraphics[width=1\textwidth]{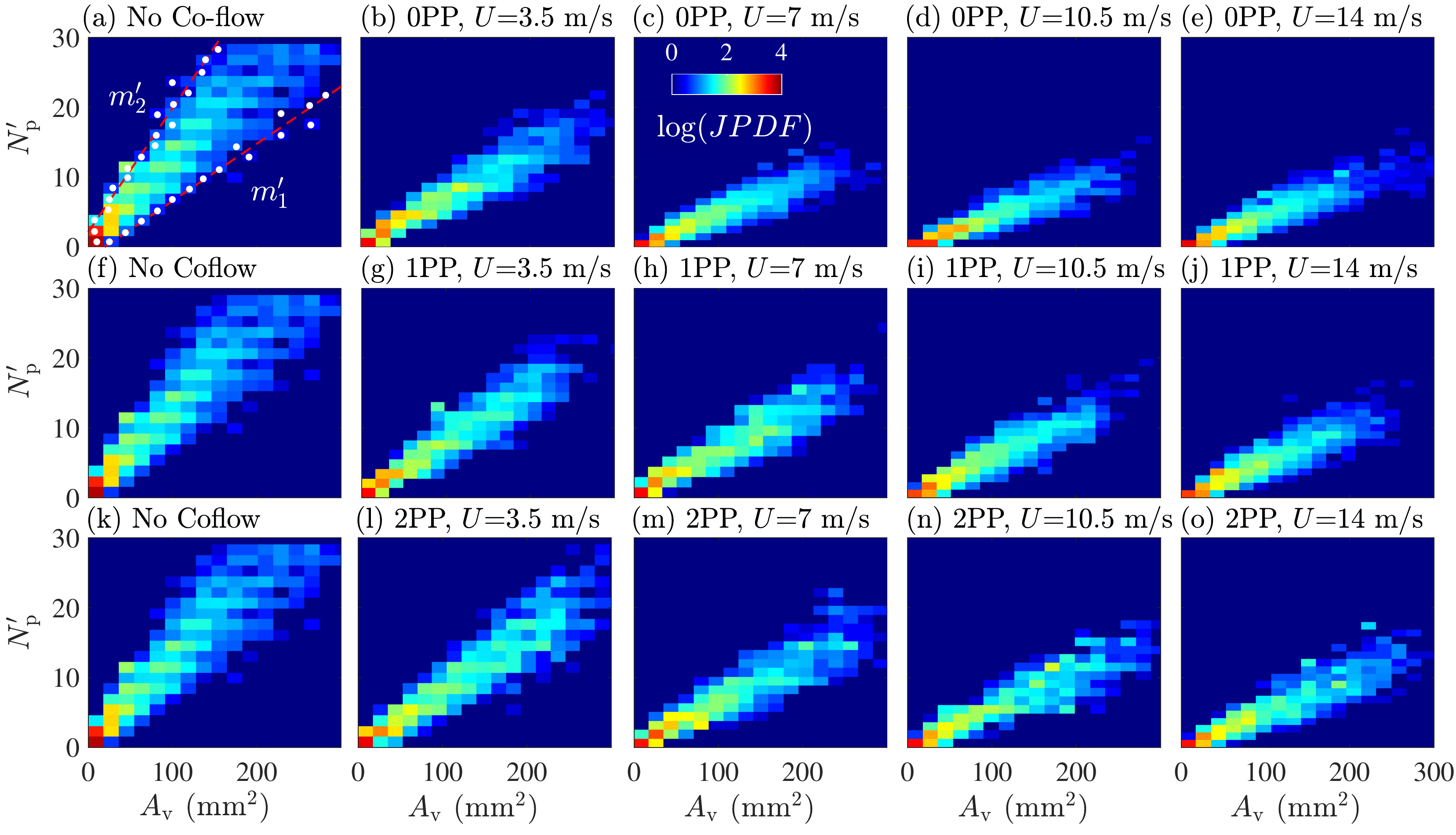}}
	\caption{The logarithmic joint probability density function of the number of droplets and void area. The first column corresponds to the no co-flow test condition and is repeating in each row for comparison purposes. The second to fifth columns correspond to the mean bulk flow velocities of 3.5, 7.0, 10.5, and 14.0~m/s, respectively. (b--e), (g--j), and (l--o) pertain to the first to third turbulence generation mechanisms, respectively.}
	\label{Fig: Void number density contour}	
\end{figure}

Figure~\ref{Fig: Number density contour} presents the JPDF of the number of droplets ($N_\mathrm{P}$) inside the clusters and the area of the clusters. The contours are presented in a logarithmic scale (with a base of 10) for improving the clarity of presentation. The JPDFs in the first column correspond to the no co-flow test condition, are repeating, and are presented for comparison purposes. The results in the second to fifth columns pertain to the mean bulk flow velocities of 3.5, 7.0, 10.5, and 14.0~m/s, respectively. The results in (b--e), (g--j), and (l--o) correspond to zero, one, and two perforated plates, respectively. As can be seen, there exists a positive correlation between the number of particles and the area of the clusters, i.e. larger clusters contain more droplets. It can also be seen that increasing the mean bulk flow velocity decreases the slope of the relation between $N_\mathrm{P}$ and $A_\mathrm{c}$, suggesting that the clusters dilute with increasing $U$. In order to quantify this, the combination of ($N_\mathrm{P},A_\mathrm{c}$) data points at which the JPDF significantly changes by changing $A_\mathrm{c}$ at a fixed $N_\mathrm{P}$ were obtained, with representative results shown by the white circular data symbols in Fig.~\ref{Fig: Number density contour}(a). The linear fits to these white data points were obtained, and the slopes of the lines were referred to as $m_1$ and $m_2$. Variations of $m_1$, $m_2$, and their average, $\overline{m} = 0.5(m_1+m_2)$, versus the mean bulk flow velocity are shown in Figs.~\ref{Fig: number density vs u}(a--c), respectively. These parameters quantify the number density of the droplets within the clusters. Similar to the above analysis, the JPDF of the number of droplets inside the voids and the area of the voids were obtained, with the corresponding results presented in Fig.~\ref{Fig: Void number density contour}. The figure indicators in Fig.~\ref{Fig: Void number density contour} are identical to those in Fig.~\ref{Fig: Number density contour}. Similar to the analysis presented for the results in Fig.~\ref{Fig: Number density contour}(a--c), the slopes ($m^\prime_1$, $m^\prime_2$, and $\overline{m^\prime} = 0.5(m^\prime_1+m^\prime_2)$) for the results in Fig.~\ref{Fig: Void number density contour} were extracted and are shown in Fig.~\ref{Fig: number density vs u}(d--f). Comparison of the results presented in Fig.~\ref{Fig: number density vs u}(a--c) with those in Fig.~\ref{Fig: number density vs u}(d--f) suggests that the number density of droplets within the clusters are generally one order of magnitude larger than that in the voids.

\begin{figure}	\centerline{\includegraphics[width=0.85\textwidth]{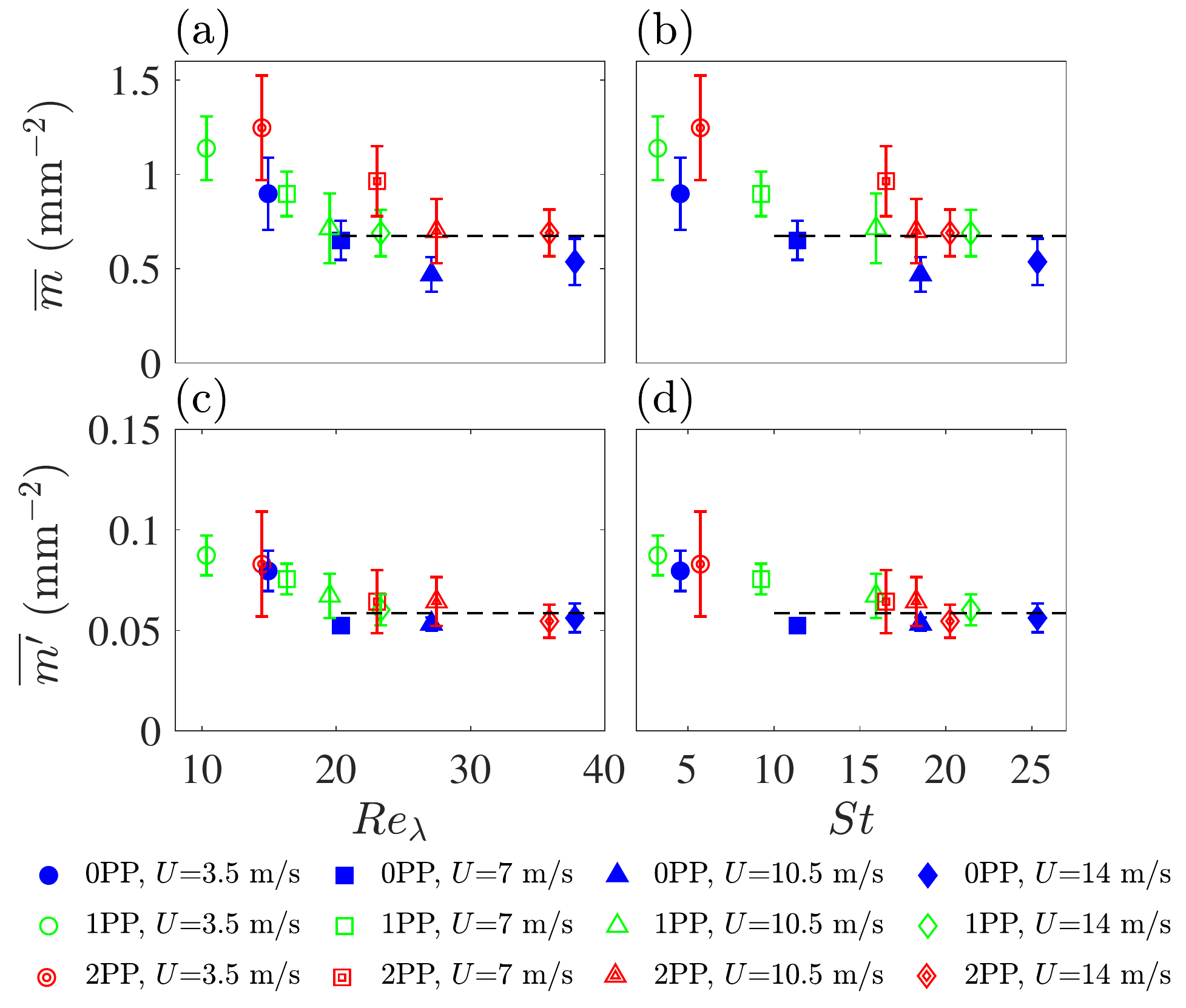}}
	\caption{(a) and (b) are variations of $\overline{m}$ versus $Re_\lambda$ and $St$, respectively. (c) and (d) are variations of $\overline{m^\prime}$ versus $Re_\lambda$ and $St$, respectively.}
	\label{Fig: number density vs st}	
\end{figure}

To investigate the effect of the non-dimensional parameters on the above number densities, the variations of $\overline{m}$ versus the Reynolds and Stokes numbers are presented in Figs.~\ref{Fig: number density vs st}(a) and (b), respectively. Also overlaid on the figures are the error bars, whose lengths correspond to $m_2-m_1$. Similarly, the variations of $\overline{m^\prime}$ versus the Reynolds and Stokes numbers are also shown in Figs.~\ref{Fig: number density vs st}(c) and (d), respectively, with the lengths of the error bars being $m^\prime_2-m^\prime_1$. The results suggest that, for $St \gtrsim 10$, the number densities of the droplets inside the clusters and voids are nearly independent of the Stokes and Reynolds numbers and equal 0.67 and 0.05~$\mathrm{mm^{-2}}$, respectively. However, for $St \lesssim 10$, increasing the Reynolds number (which is accompanied by increase of the Stokes number) nearly decreases the number densities of the droplets in the clusters and voids. 

\begin{figure}	\centerline{\includegraphics[width=0.85\textwidth]{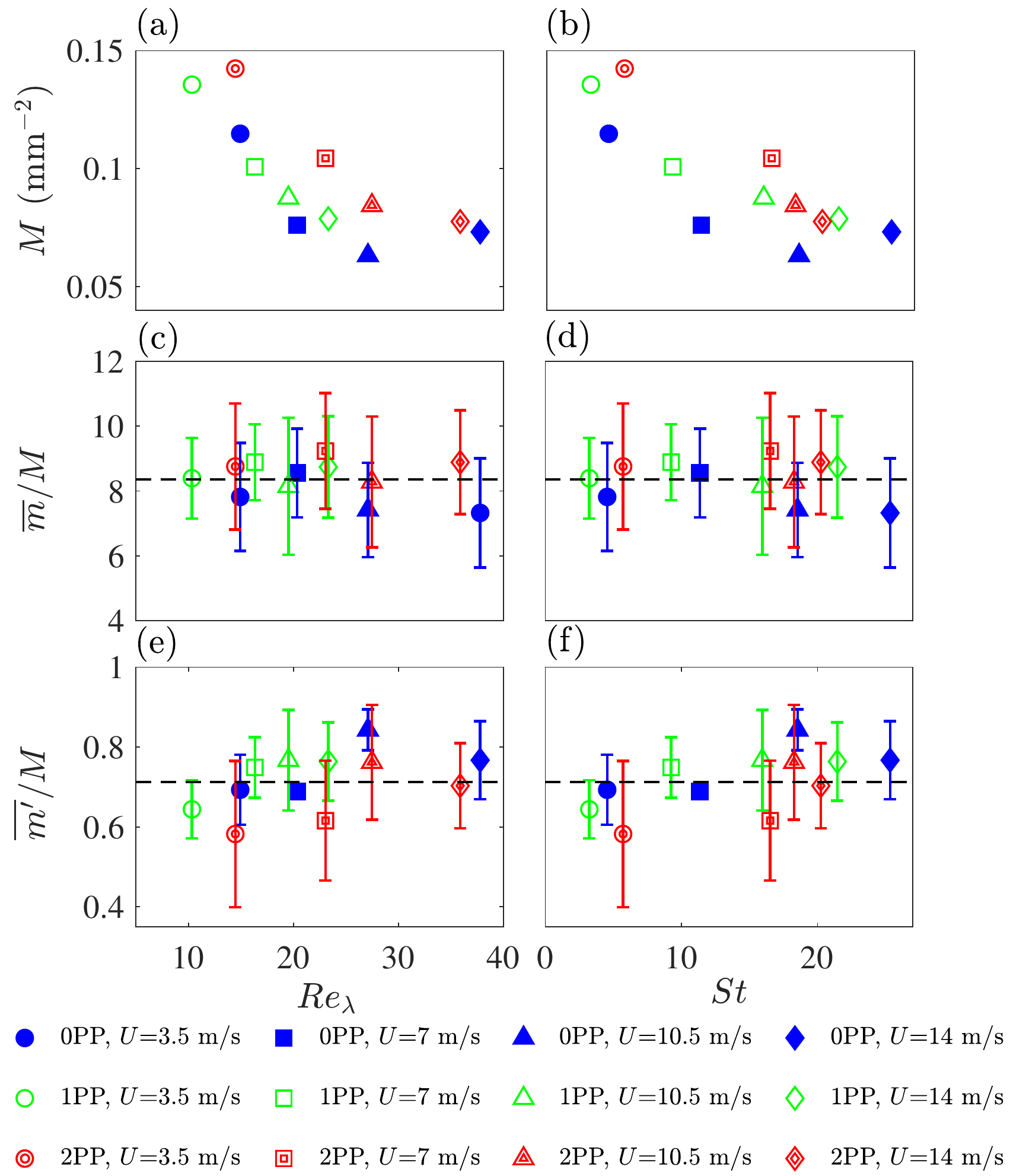}}
	\caption{(a) and (b) are the variations of the total number density versus $Re_\lambda$ and $St$, respectively. (c) and (d) are the variations of the number density of droplets in the clusters divided by the total number density versus $Re_\lambda$ and $St$, respectively. (e) and (f) are the variations of the number density of droplets in the voids divided by the total number density versus $Re_\lambda$ and $St$, respectively.}
	\label{Fig: number density ratio}	
\end{figure}

On one hand, the results presented in Fig.~\ref{Fig: tested condition} showed that increasing the Reynolds and Stokes numbers both decrease the spray volume fraction. On the other hand, for $St \lesssim 10$, the results in Fig.~\ref{Fig: number density vs st} showed that increasing both $Re_\lambda$ and $St$ also decrease $\overline{m}$ and $\overline{m^\prime}$. It is of interest to investigate the contributions of the decay of the number densities of the droplets inside the clusters and voids to the potential decay in the total number density of the droplets and if these contributions change by changing $Re_\lambda$ and $St$. Figure~\ref{Fig: number density ratio}(a)~and~(b) present the ratio of the total number of droplets detected in the Mie scattering field of view divided by the area of the field of view, $M$, versus the Reynolds and Stokes numbers, respectively. As can be seen, increasing these non-dimensional parameters decrease $M$, which is anticipated considering the decreasing trend in Fig.~\ref{Fig: tested condition}(c) and that the most probable droplet diameter is nearly unchanged by changing the test conditions. The variations of $\overline{m}/M$ versus $Re_\lambda$ and $St$ are presented in Figs.~\ref{Fig: number density ratio}(c) and (d). Similarly, the variations of $\overline{m^\prime}/M$ versus $Re_\lambda$ and $St$ are presented in Figs.~\ref{Fig: number density ratio}(e) and (f), respectively. In Figs.~\ref{Fig: number density ratio}(c and d)~and~(e and f), the lengths of the error bars are $m_2-m_1$ and $m_2^\prime-m_1^\prime$ normalized by the total droplets number density of the corresponding test condition. The results show that the number density of the droplets within the clusters and voids are 8.3 and 0.7 times the total number density, see the dashed lines in the figures, and these ratios do not change by changing the test conditions.

\subsubsection{Joint probability density function of droplet diameter and cluster/void area}
\label{subsubsec:JPDFdandA}
\begin{figure}
	\centerline{\includegraphics[width=1\textwidth]{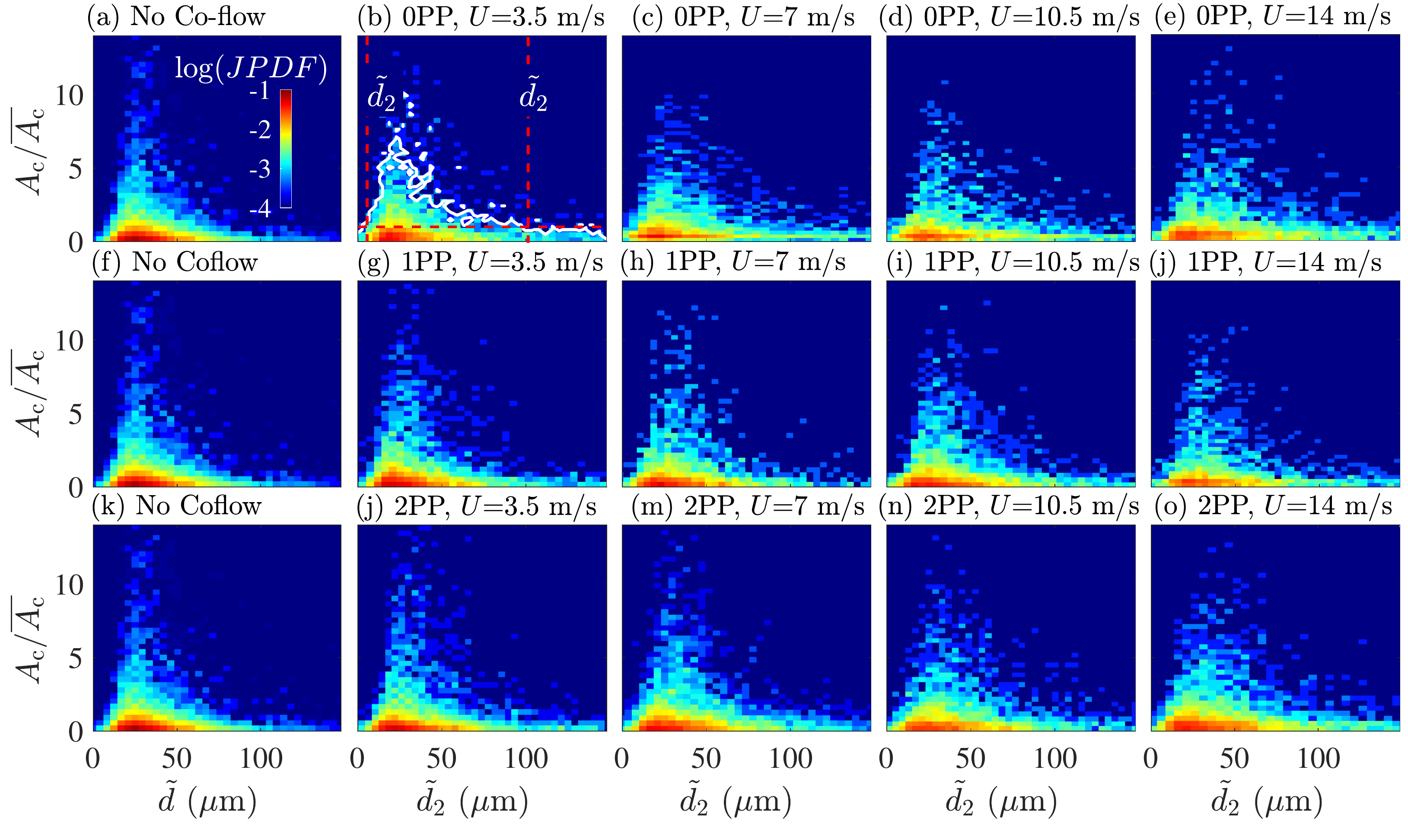}}
	\caption{The logarithmic joint probability density function of the normalized clusters area ($A_\mathrm{c}/\overline{A_\mathrm{c}}$) and the mean diameter ($\tilde{d}$) of the droplets that exist within the cluster with normalized area of $A_\mathrm{c}/\overline{A_\mathrm{c}}$. The first column corresponds to the no co-flow test condition and is repeating in each row for comparison purposes. The second to fifth columns correspond to the mean bulk flow velocities of 3.5, 7.0, 10.5, and 14.0~m/s, respectively. (b--e), (g--j), and (l--o) present the results for the first to third turbulence generation mechanisms, respectively.}
	\label{Fig: JPDF of cluster}
\end{figure}

The joint probability density function of the clusters normalized area ($A_\mathrm{c}/\overline{A_\mathrm{c}}$) and the mean diameter ($\tilde{d}$) of the droplets within the clusters with normalized area of $A_\mathrm{c}/\overline{A_\mathrm{c}}$ is presented in Fig.~\ref{Fig: JPDF of cluster}. For presentation purposes, the JPDF contours are shown in a logarithmic scale with the base of 10. The results presented in the first column pertain to the no co-flow test condition, are identical, and are shown for comparison purposes. The contours presented in the second to fifth columns correspond to the mean bulk flow velocities of 3.5, 7.0, 10.5, and 14.0~m/s, respectively. The results in Figs.~\ref{Fig: JPDF of cluster}(b--e), (g--j), and (l--o) pertain to test conditions with zero, one, and two perforated plates, respectively. As can be seen, for all test conditions, the clusters with relatively small areas ($A_\mathrm{c}/\overline{A_\mathrm{c}} \lesssim 1$) are highly likely to exist. This is due to the PDF of the cluster length scale being skewed towards small values, as shown in Fig.~\ref{Fig: PDF length scale of clusters}. It can also be seen that near $\tilde{d} \approx 30~\mu$m, which is close to the most probable droplet diameter measured in the ILIDS field of view and tabulated in Table~\ref{tab:tested onditions}, the JPDFs feature significant values for a large range of cluster sizes. This means that, for all test conditions, the majority of the clusters carry the droplets with the most probable diameter.

\begin{figure}
	\centerline{\includegraphics[width=0.8\textwidth]{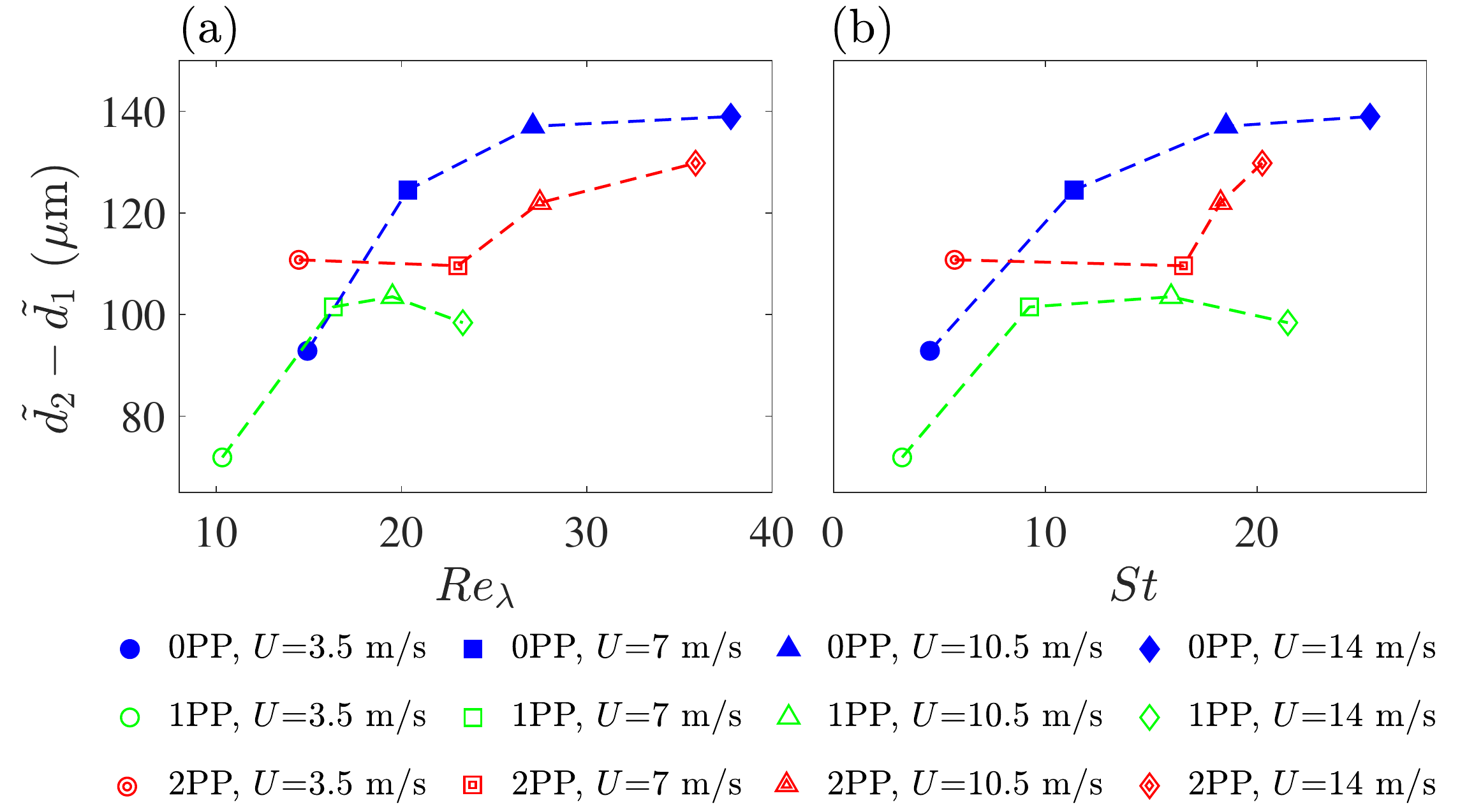}}
	\caption{(a) and (b) are the variations of $\tilde{d_2}-\tilde{d_1}$ versus  the Reynolds and Stokes numbers, respectively.}
	\label{Fig: Daimeter variability vs St}
\end{figure}

Comparison of the JPDFs for the no co-flow test condition (see the first column) with those that the co-flow was provided shows that the probability of finding large droplets (mean diameter larger than about 100~$\mu$m) within a given normalized cluster area increases by adding the co-flow. This means that the presence of the co-flow facilitates the clustering of the large droplets. In order to quantify the effect of the co-flow on $\tilde{d}$, the contour of $JPDF(A_\mathrm{c}/\overline{A_\mathrm{c}}) = 10^{-3.5}$ was considered. Then, the intersections of this contour with a given normalized cluster area (here, $A_\mathrm{c}/\overline{A_\mathrm{c}}=1$) were obtained, which are referred to as $\tilde{d_1}$ and $\tilde{d_2}$, see Fig.~\ref{Fig: JPDF of cluster}(b). The values of $\tilde{d_1}$ and $\tilde{d_2}$ change by varying the values of the selected $JPDF$ as well as the selected $A_\mathrm{c}/\overline{A_\mathrm{c}}$; however, the trends of variation for $\tilde{d_2}-\tilde{d_1}$ with the governing parameters are independent of the selected $JPDF$ value and $A_\mathrm{c}/\overline{A_\mathrm{c}}$. For all test conditions with the co-flow, the variations of $\tilde{d_2}-\tilde{d_1}$ versus $Re_\lambda$ and $St$ were obtained and presented in Fig.~\ref{Fig: Daimeter variability vs St}(a)~and~(b), respectively. The results show that increasing $Re_\lambda$ and $St$ generally increases $\tilde{d_2}-\tilde{d_1}$. As shown in Fig.~\ref{Fig: mean clusters and voids length}(d) and as tabulated in Table \ref{tab:tested onditions}, increasing the mean bulk flow velocity increases the integral length scale and decreases the Kolmogorov length scale. Thus, adding the co-flow (which is equivalent to increasing the $Re_\lambda$) increases the range of the turbulent eddy sizes. Following the mechanisms proposed in for example \cite{goto2006self}, since turbulent eddies are responsible for the clustering of the droplets, the larger range of turbulent eddy sizes (see Figs.~\ref{Fig: mean clusters and voids length}(c and d)) could increase the possibility of a broader range of droplet diameters to be positioned inside the clusters. As a result, a positive relation between $\tilde{d_2}-\tilde{d_1}$ and $Re_\lambda$ as well as $St$ is observed. This also suggests that the mechanism proposed in for example \cite{goto2006self} for droplets clustering may be extended to relatively large stokes numbers of about 25, which was examined in the present study.

\begin{figure}
	\centerline{\includegraphics[width=1\textwidth]{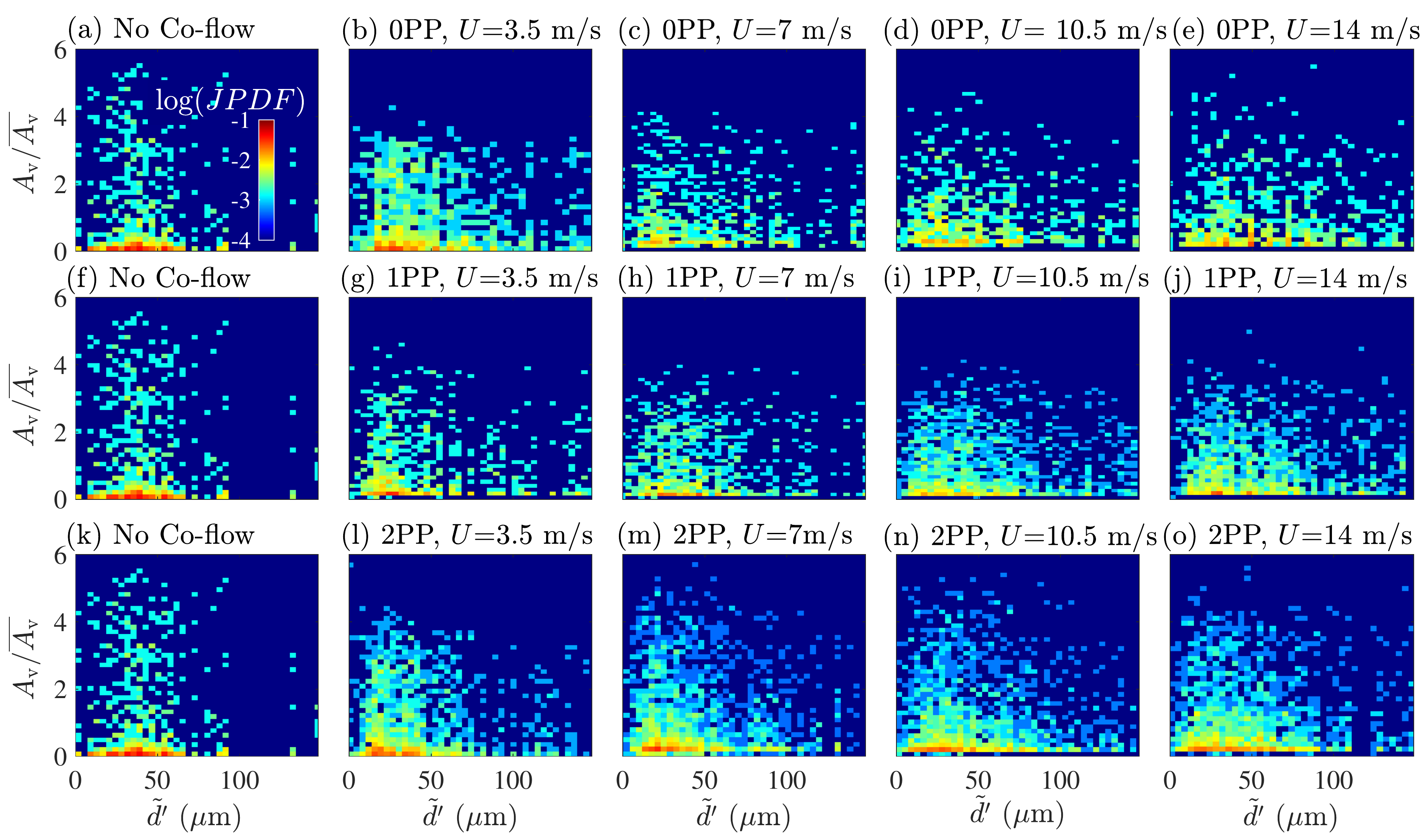}}
	\caption{The logarithmic joint probability density function of the normalized area of the voids ($A_\mathrm{v}/\overline{A_\mathrm{v}}$) and the mean diameter of the droplets that exist within the voids ($\tilde{d^\prime}$). The first column corresponds to the no co-flow test condition and is repeating in each row for comparison purposes. The second to fifth columns correspond to the mean bulk flow velocities of 3.5, 7.0, 10.5, and 14.0~m/s, respectively. (b--e), (g--j), and (l--o) present the results for the first to third turbulence generation mechanisms, respectively.}
	\label{Fig: JPDF of void}
\end{figure}

An analysis similar to the above was performed to investigate the joint characteristic of droplets and voids. Figure~\ref{Fig: JPDF of void} presents the JPDF of the voids normalized area ($A_\mathrm{v}/\overline {A_\mathrm{v}}$) versus the mean diameter of droplets within the voids ($\tilde{d^\prime}$). Similar to Fig.~\ref{Fig: JPDF of cluster}, the first to fifth columns correspond to the no co-flow, $U = 3.5$, 7.0, 10.5, and 14.0~m/s, respectively. Also, the results in the first to third columns pertain to the first to third turbulence generation mechanisms, respectively. The results presented in Fig.~\ref{Fig: JPDF of void} show that the presence of the co-flow (compared to the no co-flow condition) increases the probability of relatively large droplets to reside within the voids. The results in the second to fifth columns of Fig.~\ref{Fig: JPDF of void} suggest that increasing the mean bulk flow velocity and changing the turbulence generation mechanisms do not substantially change $JPDF(A_\mathrm{v}/\overline{A_\mathrm{v}},\tilde{d^\prime})$. This suggests that the turbulent co-flow may not facilitate positioning of the droplets within the voids.

\break
\section{Concluding remarks}
\label{Sec:Conclusions}

Separate and joint characteristics of droplets diameter and clusters/voids size at relatively large Stokes numbers were investigated experimentally. Simultaneous Mie scattering and Interferometric Laser Imaging for Droplet Sizing (ILIDS) were performed to acquire the spatial distribution of the droplets and their corresponding diameters. Also, separate hotwire anemometry was performed to characterize the background turbulent flow. A flow apparatus was utilized to produce a water spray injected in a turbulent co-flow of air. Mean bulk flow velocities of 0 (no co-flow), 3.5, 7.0, 10.5, and 14.0~m/s were examined. Three turbulence generation mechanisms (zero, one, and two perforated plates) were used in the present study. The above mean bulk flow velocities and turbulence generation mechanisms allowed to vary the Kolmogorov time scale-based Stokes number and Taylor length scale-based Reynolds number from 3 to 25 as well as 10 to 38, which correspond to relatively large and moderate values (compared to those reported in the literature), respectively. The volume fraction of the spray was varied from about $10^{-6}$ to $2\times10^{-6}$, which rendered the tested sprays as dilute.

The Vorono\"{i} analysis was performed to calculate the degree of clustering as well as to identify the clusters and voids. The results showed that, for the test conditions that one perforated plate was used, the addition of the co-flow to the spray increases the degree of clustering. However, for test conditions with one and two perforated plate(s), the addition of the co-flow decreases the degree of clustering. The results of the present study and those of past investigations were compiled, and it was obtained that the degree of clustering is primarily influenced by the Stokes number. Specifically, increasing this number from zero to about 5 increases the degree of clustering from zero to about 1.2; however, for test conditions with Stokes numbers larger than about 5, the degree of clustering does not depend remarkably on the Stokes number and plateaus at about 0.7--1. 

The clusters centers of areas were obtained and the corresponding  Vorono\"{i} cells were formed for all test conditions. The results showed that the probability density function of these cells area normalized by their corresponding mean collapse on that of the Random Poisson Process; and, as a result, the clustering of the clusters does not occur for conditions tested here. The mean locations of the clusters and voids were obtained, and it was shown that while the former is located near the core of the spray, the latter is positioned at periphery of the spray. It was shown that, at a fixed Taylor length scale-based Reynolds number, increasing the Stokes number stretches the clusters location along the jet centerline, facilitating their existence at farther distances from the spray. However, at a fixed Stokes number, increasing the Taylor length scale-based Reynolds number shrinks the region where clusters exist along the jet centerline. 

The length scales of the clusters and voids were estimated and compared with those of the background turbulent co-flow. The results showed that the mean void length scale was on the order of the integral length scale for all test conditions. However, the mean cluster length scale is smaller than the integral length scale but larger than the Taylor length scale. It was shown that increasing the Stokes number increases the mean cluster length scale to about 60 times the Kolmogorov length scale.

The results showed that the number density of the droplets within the clusters is about one order of magnitude larger than that for the voids, and both number densities decrease and then plateau with increasing the Stokes and Taylor length scale-based Reynolds numbers. It was obtained that the ratios of the clusters and voids number densities to the total number density of the spray are independent of the test conditions and nearly equal 8.3 and 0.7, respectively. The joint probability density function analysis was utilized to study the joint characteristics of the droplets and clusters/voids. It was concluded that a relatively wide range of cluster length scales can accommodate the most probable droplet diameter (about 30~$\mu$m) for all test conditions. It was shown that, although the joint PDF of the droplets diameter was not noticeably sensitive to the co-flow, increasing the Stokes and Taylor length scale-based Reynolds numbers increase the possibility of residing large droplets within the clusters. This was explained to be linked to increased range of eddy sizes interacting with droplets due to the increase of the above non-dimensional parameters.  

\section*{Acknowledgments}
The authors are grateful for the financial support from The University of British Columbia through the Eminence Fund program. Sina Kheirkhah acknowledges the financial support from the Canada Foundation for Innovation.

\section*{Declaration of interest}
The authors report no conflict of interest.

\section*{Appendix A: Spray volume flow rate calibration}
\begin{figure}
	\centerline{\includegraphics[width=0.5\textwidth]{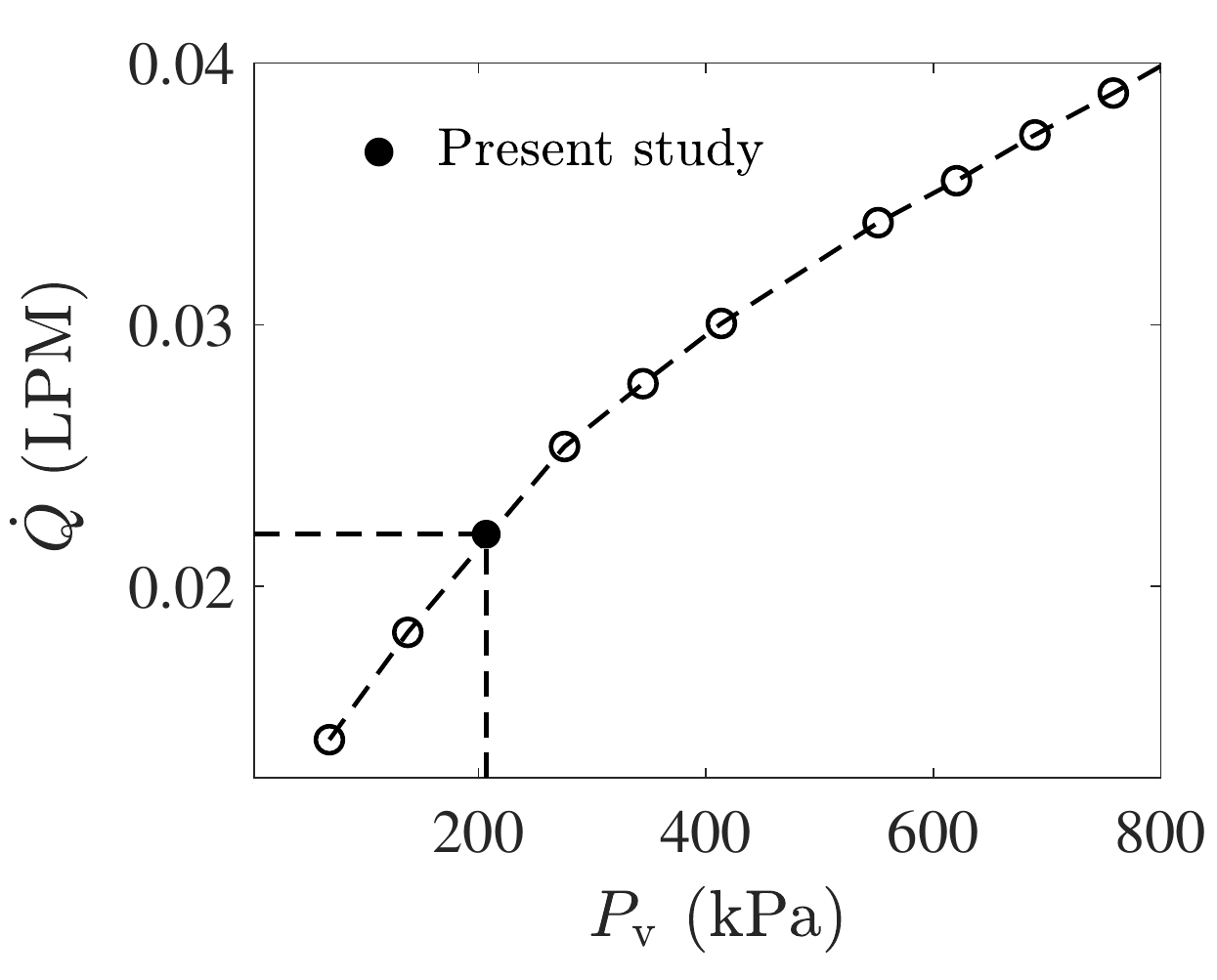}}
	\caption{Calibration relation between the water vessel pressure and the spray flow rate.}
	\label{Fig: Nozzle calibration curve}
\end{figure}

For the spray volume flow rate ($\dot{Q}$) calibration, first, the spray injector shown in Fig.~\ref{Fig:Nozzle geometry} was connected to a sealed container; and then, the upstream pressure of the injector ($P_\mathrm{v}$) was set to several values ranging from about 50 to 800~kPa using the pressure controller shown as item (2) in Fig.~\ref{Fig:Setup}. For each set value of the upstream pressure, a high-precision scale was used to measure the collected water mass, which was converted to volume by using the water density at the laboratory temperature of 20$^\mathrm{o}$C. Then, the volume flow rate was calculated by dividing the volume of the collected water (in liters) to the time duration of the calibration experiment, which was 120~s. Figure~\ref{Fig: Nozzle calibration curve} presents the variation of $\dot{Q}$ versus $P_\mathrm{v}$. For all conditions tested in the present study, $P_\mathrm{v}$ was set to 206~kPa, which led to a spray volume flow rate of 22 cubic centimeter per minute (see the black circular data symbol in Fig.~\ref{Fig: Nozzle calibration curve}).   
\label{appA}

\section*{Appendix B: Procedure for registering the Mie scattering images to the ILIDS images}
\label{appB}

The centers of the droplets identified from the ILIDS images are not identical to those obtained form the Mie Scattering images, and a discrepancy exists between the centers of the droplets identified from the above diagnostics. In addition to the present study, such discrepancy is also noted and discussed in the past investigations, see for example \cite{boddapati2020novel} and \cite{hardalupas2010simultaneous}. For the droplets and clusters/voids joint characteristics calculations, it is important to remove the discrepancy between the centers of the droplets obtained from the Mie scattering and ILIDS images. The procedure for removing the above discrepancy and thus registering the Mie scattering to the ILIDS images is discussed in the following. 

\begin{figure}
	\centerline{\includegraphics[width=0.9\textwidth]{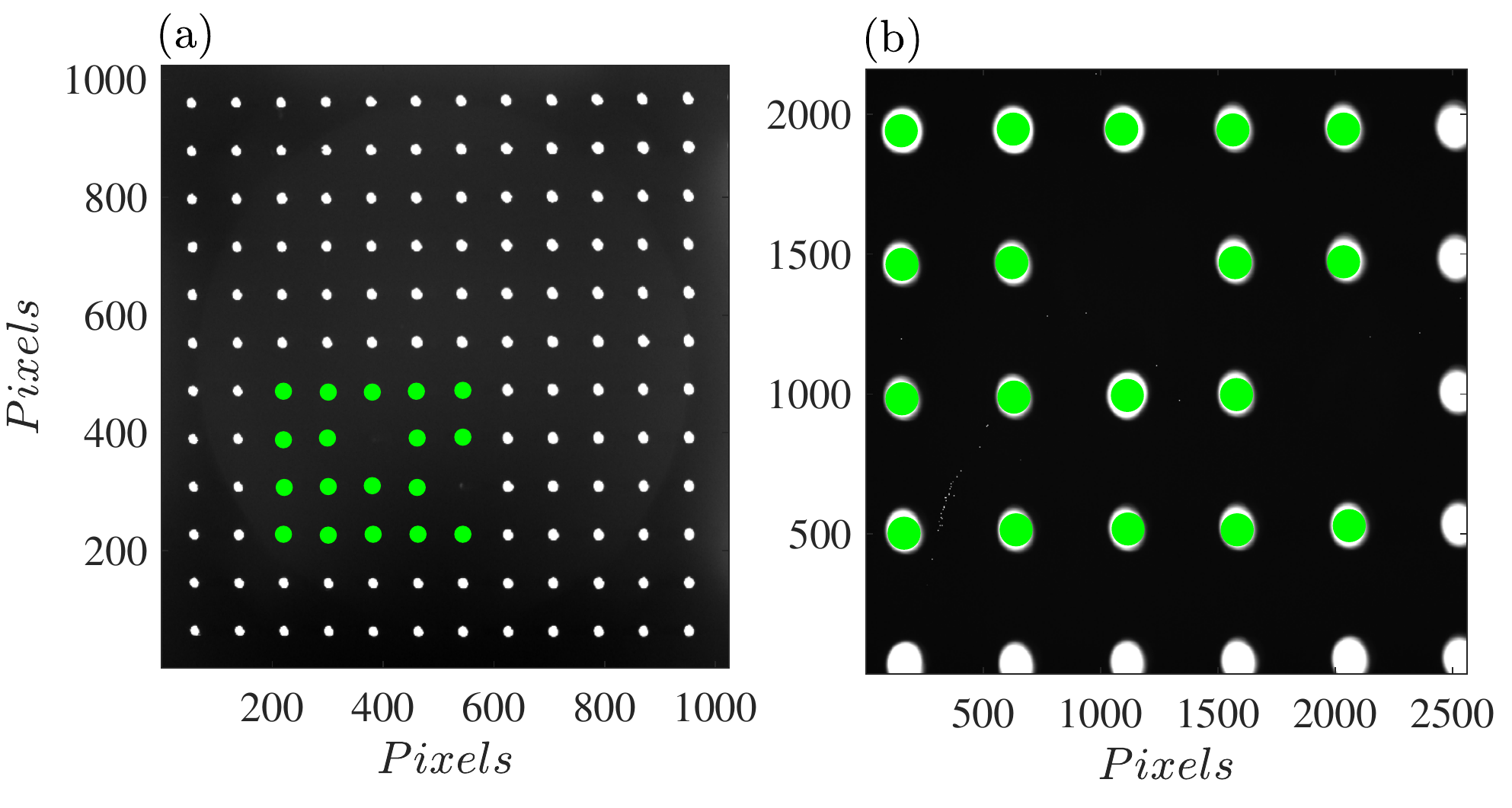}}
	\caption{(a) and (b) are the images of the target plate captured by the Mie Scattering and ILIDS cameras, respectively.}
	\label{Fig:Target plate}
\end{figure}

\begin{figure}
	\centerline{\includegraphics[width=1\textwidth]{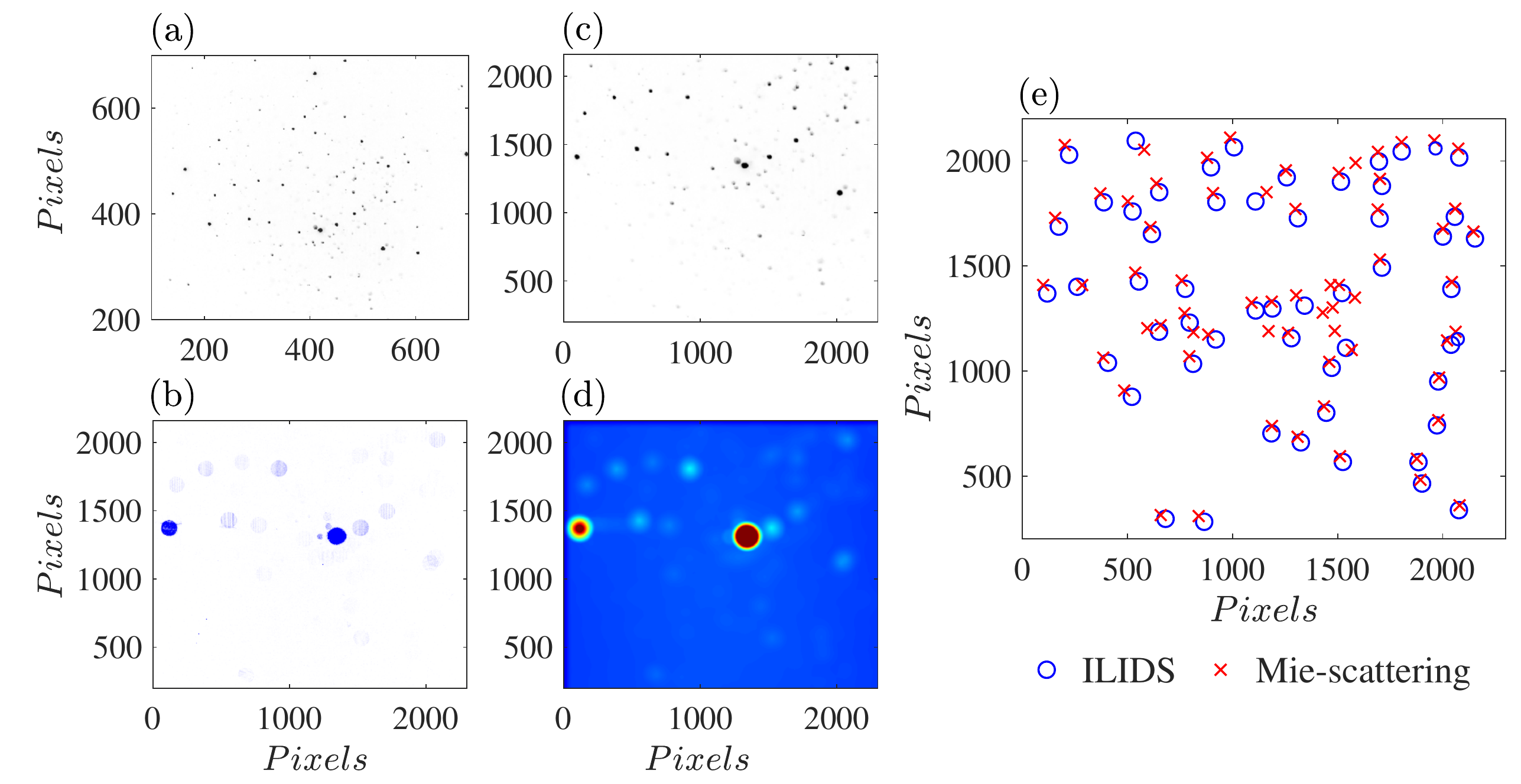}}
	\caption{(a) and (b) are representative raw Mie scattering and ILIDS images, respectively. (c) is the image obtained by mapping that shown in (a) to the imaging plane of the ILIDS camera. (d) is the convolution of the results shown in (b) with a disk-shaped mask. (e) presents the centers of the droplets obtained from the ILIDS and Mie scattering images.}
	\label{Fig: Image registration}
\end{figure}

First, a 2D target plate was manufactured and positioned in the location illuminated by the laser sheet in the experiments. Then, the images of the target plate were captured by the cameras used for collecting the Mie scattering and ILIDS images, with the collected images shown in Figs.~\ref{Fig:Target plate}(a) and (b), respectively. Eighteen points, see the green circles in Fig.~\ref{Fig:Target plate}(a), were considered and were identified in Fig.~\ref{Fig:Target plate}(b). The collective geometrical shape of these points was purposely selected to be asymmetric, facilitating the above identification. Then, the matrix that transforms data points in Fig.~\ref{Fig:Target plate}(a) to those in Fig.~\ref{Fig:Target plate}(b) was obtained. This transformation matrix was used for registering the Mie scattering images to the ILIDS images, with a representative result for such transformation discussed below. 

A representative and simultaneously acquired Mie scattering and ILIDS images are shown in Figs.~\ref{Fig: Image registration}(a) and (b), respectively. The results correspond to the test condition with one perforated plate and the mean bulk flow velocity of 10.5~m/s. Figure~\ref{Fig: Image registration}(c) presents the results in Fig.~\ref{Fig: Image registration}(a) after the application of the above transformation. Figure~\ref{Fig: Image registration}(d) presents that in Fig.~\ref{Fig: Image registration}(b) after application of the convolution discussed in subsection~\ref{subsec:ILIDSdatareduction}. The procedure discussed in subsection~\ref{subsec:ILIDSdatareduction} was followed  to identify the centers of the droplets from the results shown in Fig.~\ref{Fig: Image registration}(d), with the corresponding centers shown by the blue circular data points in Fig.~\ref{Fig: Image registration}(e). Overlaid on Fig.~\ref{Fig: Image registration}(e) are also the centers of the droplets obtained from binarizing the Mie scattering image in Fig.~\ref{Fig: Image registration}(c). The centers of the droplets obtained from the Mie scattering image are shown by the red cross data symbols in Fig.~\ref{Fig: Image registration}(e). As can be seen, the spatial locations of the blue circular data symbol are close to those of the red cross data symbol. Thus, for calculating the joint characteristics of the droplets and clusters (voids), the centers identified from the ILIDS image were replaced by those obtained from the Mie scattering image.             

\bibliography{Bib}
\bibliographystyle{jfm}

\end{document}